\documentclass[acmsmall]{acmart}

\AtBeginDocument{%
  \providecommand\BibTeX{{%
    \normalfont B\kern-0.5em{\scshape i\kern-0.25em b}\kern-0.8em\TeX}}}

\setcopyright{acmcopyright}
\copyrightyear{2018}
\acmYear{2018}
\acmDOI{XXXXXXX.XXXXXXX}

\acmJournal{PACMHCI}
\acmYear{2024} \acmVolume{x} \acmNumber{CSCW}
\acmArticle{xxx} \acmMonth{11}
\acmPrice{}\acmDOI{xx.xxxx/xxxxxxx} 

\usepackage{multirow}
\usepackage{soul}
\usepackage{dirtytalk}

\usepackage{subcaption}
\usepackage{enumitem}
\usepackage{multirow}
\usepackage{color}

\newcommand{\comments}{1}
\newcommand{\ignore}[1]{}
\ifdefined\comments

\else
\fi

\begin{document}

\title[]{Does More Advice Help? The Effects of Second Opinions in AI-Assisted Decision Making}









\author{Zhuoran Lu}
\affiliation{%
  \institution{Purdue University}
  \city{West Lafayette}
  \country{USA}}
\email{lu800@purdue.edu}

\author{Dakuo Wang}
\affiliation{%
  \institution{Northeastern University}
  \city{Boston}
  \country{USA}}
\email{mingyin@purdue.edu}

\author{Ming Yin}
\affiliation{%
  \institution{Purdue University}
  \city{West Lafayette}
  \country{USA}}
\email{mingyin@purdue.edu}

\begin{abstract}
AI assistance in decision-making has become popular, yet people's inappropriate reliance on AI often leads to unsatisfactory human-AI collaboration performance. In this paper, through three pre-registered, randomized human subject experiments, we explore whether and how the provision of {\em second opinions} may affect decision-makers' behavior and performance in AI-assisted decision-making. 
We find that if both the AI model's decision recommendation and a second opinion are always presented together, decision-makers reduce their over-reliance on AI while increase their under-reliance on AI, regardless whether the second opinion is generated by a peer or another AI model. However, if decision-makers have the control to decide when to solicit a peer's second opinion, we find that their active solicitations of second opinions have the potential to mitigate over-reliance on AI without inducing increased under-reliance in some cases.
We conclude by discussing the implications of our findings for promoting effective human-AI collaborations in decision-making. 
\end{abstract}





\maketitle

\section{Introduction}
With its rapid development in recent years, Artificial Intelligence (AI) technology has been integrated into many industries,  such as business~\cite{hui2000data,sabbeh2018machine,wang2019human,jahanbakhsh2023exploring,wang2022documentation}, healthcare~\cite{kohli2018application,kourou2015machine,wang2023human,zhang2023rethinking}, education~\cite{zhang2022storybuddy}, transportation~\cite{nguyen2018deep}, and more. A common way for AI to augment human workflows in various domains is through \textbf{{\em AI-assisted decision-making}}, that is, an AI-based decision aid provides decision recommendations to humans while humans make the final decisions. As humans and AI may each possess unique intelligence that is complementary to each other, the \textbf{\em{human-AI collaboration}}~\cite{wang2020human} in this decision-making scenario has the potential to utilize the best of humans and AI and realizes a joint performance beyond what can be achieved by each party alone.

In reality, however, the joint decision-making performance of the human-AI team is often not as good as expected. One primary reason underlying such unsatisfactory human-AI collaboration is that humans often rely on the AI recommendations {\em inappropriately}. Humans may not trust an AI model and hence avoid adopting its recommendations even when the recommendations are highly accurate, resulting in \textbf{{\em under-reliance}} on AI~\cite{dietvorst2015algorithm}. On the other hand, sometimes humans also show a degree of \textbf{{\em over-reliance}} on AI, as they blindly accept the recommendations of an AI model even when it makes sizable mistakes~\cite{poursabzi2021manipulating,buccinca2021trust,chiang2021you}. To help people establish a more appropriate level of reliance on AI and improve the human-AI joint decision-making performance, researchers and practitioners have explored a wide range of methods, such as enhancing humans' understandings of the rationale underlying AI recommendations~\cite{bansal2021does,schemmer2022should,yang2020visual,schemmer2022meta,yao2023beyond}, enforcing people to engage in careful deliberation~\cite{buccinca2021trust,park2019slow}, and communicating to people the importance of their decisions~\cite{ashktorab2021ai}. However, mixed results have been reported regarding the effectiveness of these methods.

This challenge of humans inappropriately relying on suggestions provided by some ``advisors'' is not new. Indeed, in the classical paradigm of ``Judge-Advisor System'' in the advice taking research~\cite{bonaccio2006advice}, where a human ``judge'' receives suggestions from another human ``advisor'' before making their final judgement on a decision-making problem, it is also observed that the judge may inappropriately discount the advisor's suggestions~\cite{yaniv2000advice} or over-utilize the advisor's low-quality advice~\cite{schultze2017inability}. 
Interestingly, a common intervention adopted in these scenarios to improve the human judge's decision-making quality is to introduce advice from a second advisor, so that the judge can explore different perspectives and suggestions to make a better final decision~\cite{yaniv2004benefit,yaniv2007using,budescu2000confidence,johnson2001averaging}. Naturally, one may wonder if similar methods will also benefit AI-assisted decision-making---If second opinions from other {\em human peers} are presented to a decision-maker in addition to the AI recommendation, 
can they help the decision maker rely on AI more appropriately and achieve a higher level of decision-making performance?
As a motivating example,
consider an investor who is assisted by an AI model in deciding their stock trading strategies~\cite{patel2015predicting,kumbure2022machine,vijh2020stock,xie2022word}:
when the investor is about to buy/sell a stock given an AI model's recommendation, will the presence of a second opinion from other investors (e.g., from online discussion forum \textit{wallstreetbets} in reddit~\cite{betzer2022online,long2022just}) help them make better investment decisions (or the opposite)? Thus, our first goal in this study is to answer the following question:
\begin{itemize}
        \item \textbf{RQ1}: How do second opinions from {\em human peers} affect decision-makers' reliance on the AI model (e.g., over-reliance, under-reliance, and appropriate reliance) and influence their decision-making performance (e.g., decision accuracy, time and confidence)?
\end{itemize}

There are reasons to conjecture the answer to this question either way. On the one hand, it is possible that the decision-maker (e.g., the investor) may perceive the AI model to be more competent than their human peers (e.g., AI has the ``expert power'' or authority) ~\cite{hou2021expert,kapania2022because}; if so, the presence of second opinions from peers may hardly change how they interact with the AI model. On the other hand, after observing potential disagreements between the AI model and the peers on some decision-making cases, decision-makers may evaluate
the AI recommendations more critically and incorporate 
them into their final decisions more intelligently, which may result in an improvement in their decision-making accuracy. In this sense, perhaps second opinions from those who oppose the AI more frequently can lead to a larger accuracy improvement~\cite{de2022doubting}. 
Another possibility is that the decision-maker may leverage the level of agreement between the AI recommendations and the second opinions from peers as a heuristic to gauge the trustworthiness of the AI model and adjust their reliance strategies accordingly. However, how changes in the decision-maker's reliance on AI translate to changes in their decision-making accuracy is not clear in this case. 
Finally, we note that beyond decision accuracy, decision-making performance can also be evaluated by other metrics, such as how efficiently the decisions are made (e.g., decision time) and the degree that the decision-makers' subjective perceptions of their decisions (e.g., decision confidence) are calibrated; it's thus necessary to examine how the provision of second opinions from human peers will affect these aspects of performance to obtain a comprehensive understanding.  

In addition, suppose second opinions from human peers have significant impacts on decision-makers' reliance behavior and performance in AI-assisted decision making, a natural follow-up question to ask is whether these impacts are caused by the {\em content} of the second opinions or the stated {\em source} of the second opinions. As one could have solicited second opinions from another AI model instead of human peers in AI-assisted decision making, the second research question we aim to answer in our study is:

\begin{itemize}
    \item \textbf{RQ2:} 
    Do the impacts of 
    second opinions on decision-makers' reliance on AI and performance in AI-assisted decision making change, when 
    the second opinions are claimed to be solicited from {\em another AI model} rather than human peers?
\end{itemize}

To obtain a thorough understanding of these two questions, we conducted two pre-registered, randomized human-subject experiments (Experiment 1: $N = 428$, Experiment 2: $N = 516$) on Amazon Mechanical Turk (MTurk). In these experiments, subjects were asked to complete a series of sentiment analysis tasks to decide whether a movie review is positive or negative, with the decision recommendations provided by an AI model. 

Specifically, in Experiment 1, we created four treatments by varying {\em whether} second opinions generated by human peers were presented to subjects on each decision-making task, and if so, {\em how frequently} they agreed with the AI recommendations. 
This design, thus, enabled us to understand whether the effects of peer-generated second opinions on decision-makers' reliance behavior and performance in AI-assisted decision making are moderated by the level of agreement between the peers and the AI model.  
Our results showed that when second opinions from human peers are always presented to decision-makers, they result in significant decreases in decision-makers' over-reliance on the AI model but also trigger significantly increased level of under-reliance. These changes are especially salient as the peers disagree with the AI model more frequently. Overall, we do not find the presence of peer-generated second opinions significantly changes decision-makers' accuracy in AI-assisted decision making, but it does result in significant increases in decision-makers' decision time and their confidence in their correct decisions. 

For Experiment 2, we set up five treatments: a control treatment where second opinions were never presented to decision-makers, and four experimental treatments where second opinions were always presented to decision-makers on every task.  
In addition, the four experimental treatments were arranged in a 2 by 2 factorial design varying along two dimensions---{\em the frequency of agreement} between the second opinions and the AI model (i.e., low vs. high), and {\em the stated source} of the second opinions (i.e., human peers vs. another AI model). We observed similar results in this experiment as those obtained in Experiment 1, regardless of the stated source of the second opinions. This means that the impacts of second opinions on decision-makers are mainly due to the content of the second opinions rather than their sources.

Both Experiments 1 and 2 suggest that simply providing second opinions on all decision-making tasks may fall short in helping improve decision-makers' accuracy in AI-assisted decision making.  Interestingly, in the exploratory analysis of both experiments, we found that a key reason that potentially limits the benefits of providing second opinions is the frequent presence of disagreeing second opinions on tasks where the AI recommendation is {\em correct}.
This observation sparked a natural idea---Instead of always presenting second opinions, we can allow decision-makers to actively {\em request} for second opinions only when they need it. Ideally, we hope decision makers may have some capability in differentiating the correctness of AI recommendation, thereby decreasing the solicitation of second opinions on tasks where the AI recommendation is correct.   
This leads to our final research question:

\begin{itemize}
    \item \textbf{RQ3}: How does {\em having the option to solicit} second opinions affect decision-makers' reliance on the AI model and influence their performance in AI-assisted decision making?
\end{itemize}

To answer this question, we conducted a third pre-registered, randomized human-subject experiment (Experiment 3, $N = 336$) on MTurk, again having subjects complete AI-assisted sentiment analysis tasks. In this experiment, instead of presenting a second opinion on every task, we provided subjects in some treatments with the option to actively {\em request} for a second opinion if they needed it. Results we obtained from all subjects of this experiment, regardless of whether they had ever requested for any second opinions on any task, showed a similar trend as the results in the first two experiments, except that the option of soliciting second opinions no longer increases decision time. Nevertheless, when we focused on the comparisons between those subjects who had requested for second opinions {\em at least on some task} and the comparable subjects in the control treatment 
who never saw any second opinions, we found that decision-makers' active solicitations of second opinions may result in a decrease in over-reliance {\em without} inducing higher levels of under-reliance; however, this is only observed in the treatment where the level of agreement between the second opinion and the AI model's recommendation is relatively high.

Taken together, our results highlight the promise of introducing second opinions 
as an intervention in the AI-assisted decision-making workflows to help people rely on AI more appropriately and eventually improve their AI-assisted decision-making performance. Meanwhile, the effectiveness of this intervention is shown to be dependent on both the ways that the second opinions are presented and the characteristics of the second opinions. We conclude by discussing the design implications and limitations of our work.

\section{Related Work}

\subsection{AI-assisted Decision-Making} 
The increasing prevalence of AI assistance has sparked great interest in the research community to empirically understand how people interact with, trust, and rely on AI models during this collaborative decision making process~\cite{lai2021towards,wang2023human}. 
 {Early studies focus on understanding human decision-makers' preferences between decision recommendations made by humans or AI models. Human subjects in these studies were often asked to explicitly choose to receive recommendations from either humans or AI, or be presented with the same recommendation that was labeled as from either humans or AI. Interestingly, both the phenomenon of ``algorithm aversion'' (i.e., recommendations from humans are used more than those from AI)}~\cite{dietvorst2015algorithm,yeomans2019making,erlei2022s} {and ``algorithm appreciation'' (i.e., recommendations from AI models are used more than those from humans)}~\cite{logg2019algorithm} {are observed in different contexts.}
More recently, researchers start to identify a wide range of factors that can affect people's reliance on AI's decision recommendations. For example, performance indicators and feedback of the AI model, such as its accuracy~\cite{yu2019trust,yin2019understanding,lai2019human}, confidence~\cite{rechkemmer2022confidence,zhang2020effect}, and the expectation and first impressions of the model competency~\cite{kocielnik2019will,tolmeijer2021second,nourani2021anchoring,nourani2020investigating,tolmeijer2021second}, are shown to significantly impact people's willingness to rely on the AI model. When performance-related information is absent, it is found that people may utilize other heuristics or cues to determine how to rely on the AI model, including how frequently the AI recommendations align with their own judgments~\cite{lu2021human} and their mental models of the AI model's error boundaries~\cite{bansal2019beyond,bansal2019updates}.

Meanwhile, despite it is believed that human-AI collaborations in AI-assisted decision-making may enable the human-AI team to outperform either party alone in their joint decision-making performance, it is widely observed in empirical studies that achieving such human-AI complementarity is quite challenging~\cite{hemmer2021human,bansal2021does}. A key limiting factor is that in their interactions with AI models, humans often exhibit a degree of {\em inappropriate reliance} on AI. For example, people may fail to reject incorrect AI recommendations, resulting in {\em over-reliance} on AI~\cite{schoeffer2023interdependence,bansal2021does,nourani2020role,logg2019algorithm}, while there are also times that human decision-makers do not adopt highly accurate AI recommendations, resulting in {\em under-reliance} on AI~\cite{lu2021human,vaughan2020human,dietvorst2015algorithm}. 

\subsection{Approaches to Promote People's Appropriate Reliance on AI} 
In light of people's inappropriate reliance on AI in AI-assisted decision-making, in recent years, a wide range of approaches have been developed to help people rely on AI more appropriately and improve the decision accuracy of the human-AI team. For example, a simple indicator of AI confidence may help people calibrate their reliance on the AI model~\cite{mcguirl2006supporting,zhang2020effect}. Another commonly used intervention is to provide AI explanations along with the decision recommendations, which allow people to probe into the AI model's rationale before determining whether to rely on its recommendations~\cite{zhang2020effect,ehsan2022human,wang2019designing,yang2020visual,lai2020chicago,bansal2021does,liu2021understanding,cheng2019explaining,chen2023understanding}. However, the effectiveness of AI explanations in promoting appropriate reliance on AI is inconclusive~\cite{fok2023search}.
 {For instance,}  \citeauthor{carton2020feature}~\cite{carton2020feature}  {found that the feature-based explanation does not help people detect online toxic content more accurately when they are assisted by toxic text classifiers, but the same type of explanation was shown to improve people's accuracy in  AI-assisted recidivism risk assessments}~\cite{green2019principles}. 
 {Researchers also found that the effectiveness of different types of explanations on helping people calibrate their reliance on AI varies, and sometimes the provision of certain kinds of explanations may even result in significant over-reliance on AI models for some people} ~\cite{wang2021explanations,yang2020visual,chu2020visual,bansal2021does,schaffer2019can,bussone2015role}---\citeauthor{schaffer2019can}~\cite{schaffer2019can}  {showed that showing explanations to people who reported higher familiarity with the decision making tasks led to automation bias, while detailed explanations were also found to lead users to develop over-reliance on AI}~\cite{bussone2015role,passi2022overreliance}.  {A recent meta-analysis reports that people's decision making performance does not have significant differences between the cases that they are assisted by an AI model with or without the explanations}~\cite{schemmer2022meta},  {suggesting that overall, the effects of current AI explanations on promoting appropriate reliance is somewhat limited}. 

Beyond various approaches to change the ways AI recommendations are communicated, additional interventions have been designed to promote appropriate reliance on AI through influencing humans~\cite{wischnewski2023measuring, passi2022overreliance}.  
For instance, cognitive forcing functions \cite{buccinca2021trust} have been used to encourage people to engage with the AI recommendations more cognitively, which are shown to reduce people's over-reliance on AI significantly. Frameworks have been proposed to monitor people's trust and reliance on AI and use cognitive cues to prompt them to re-calibrate their trust and reliance when needed~\cite{okamura2020adaptive}. It is also found that before people start their interactions with AI models, carefully designed training can help to enhance people's AI literacy and their overall understanding of the AI model's behavior~\cite{chiang2022exploring,chiang2021you,lai2020chicago,erlei2020impact,he2023knowing,he2023knowing}, which result in a decrease in people's inappropriate reliance on the AI model during the actual interactions. Most recently, researchers have also explored the use of computational approaches to model and predict people's interactions with AI models~\cite{wang2022will,li2023modeling,li2024decoding}, which inform the designs of adaptive interfaces to improve people's appropriate reliance on AI~\cite{ma2023should}.


 {For a more comprehensive review on people's inappropriate reliance on AI (especially  over-reliance) and mitigation methods, please see{~\cite{passi2022overreliance}}. We note that  
the existing approaches for promoting people's appropriate reliance on AI are rarely panaceas---some interventions reduce over-reliance with the price of increasing under-reliance (e.g., cognitive forcing functions), while the success of other approaches (e.g., AI literacy interventions) is only observed for people with certain characteristics. Nevertheless, all of these studies contribute important insights into what may or may not work, and when they can work, when it comes to mitigating inappropriate reliance on AI.}



\subsection{Advice Taking}
\label{advice_distance}

How humans take advice from others in their decision making has been studied for decades in psychology~ 
\cite{bonaccio2006advice}. This research often adopts a particular advice structure called ``Judge–Advisor System'' (JAS) that is very similar to the structure of AI-assisted decision-making---in JAS, the advisor provides advice to the judge, while the judge is responsible for making the final decision. It is found that judges often have some capability in perceiving the quality of the advice; hence they utilize good advice more than bad advice~\cite{harvey2000using,harvey2000using,yaniv2000advice}. However, judges' utilization of advice is often not optimal. For instance, due to their egocentric bias, judges may associate a very high weight with their own opinions and, therefore, significantly discount advice that is distant from their own opinions~\cite{yaniv2000advice}.

To help further improve the decision quality of judges, advice from multiple sources is often provided to the judge so that the judge can aggregate multiple pieces of advice~\cite{yaniv2007using}. Indeed, there is empirical evidence showing that  
by aggregating the opinions from different advisors, in many cases, judges can make more accurate or even optimal decisions~\cite{harvey1997taking,yaniv2004benefit,yaniv2007using,budescu2000confidence,johnson2001averaging}. However, how the judge incorporates multiple advice into their decision, and how exactly this affects the judge's decision quality depend on many factors~\cite{schotter2003decision,soll1999intuitive,druckman2001using,gino2008blinded,hutter2016seeking}. 
For instance, one relevant factor is the degree of ``conflict'' between advisors (i.e., the level of disagreement between advisors' opinions)~\cite{rohrbaugh1979improving,sniezek1989examination,sniezek1995cueing}. In the ideal scenario, conflicts among advisors could result in improvement in the judge's decision performance because the judge's blind trust in any single advisor is decreased~\cite{ronis1987components}.
It is found that integrating multiple {\em independent} pieces of advice is particularly helpful for increasing the judge's decision accuracy gain~\cite{yaniv2004benefit,bonaccio2006advice}. These promising findings inspire us to investigate that, in AI-assisted decision-making, how the provision of second opinions generated independently by human peers or another AI model in addition to the AI recommendation will affect people's reliance on AI. More generally, we wonder how the potential agreement and disagreement between two independent advisors' opinions affect people's behavior and performance in AI-assisted decision making.

\section{Experiment 1: 
Peers' Second Opinions Always Presented
}
To understand how the presence of second opinions from human peers affect people's behavior and performance in AI-assisted decision-making, and how these effects vary with the agreement level between the peers and the AI model, we recruit human subjects from Amazon Mechanical Turk (MTurk) and conduct our first randomized experiment. 

\subsection{Experimental Task}
\label{sec:task}
\noindent \textbf{Choice of decision-making task.}  In our experiment, we asked subjects to determine the sentiment of movie reviews with the help of an AI model. 
Specifically, in each task, subjects were presented with a movie review taken from the IMDB movie review dataset~\cite{maas-EtAl:2011:ACL-HLT2011}, and the length of the review was controlled to be between 280 and 300 words.
Along with the movie review, we also showed subjects an AI model's binary prediction of the review's sentiment (i.e., positive vs. negative), while subjects in some experimental treatments also had access to the judgement of the review's sentiment made by a peer (i.e., a randomly selected crowd worker; see Section~\ref{sec:design} for details). After reviewing all this information, subjects were asked to make a final decision on whether the sentiment expressed in the movie review was positive or negative. In total, each subject needed to review the same set of 20 movie reviews in our experiment. 



We used the sentiment analysis task in our experiment for several reasons. To begin with, accurately analyzing sentiment is crucial in various industries, including retail~\cite{fiarni2016sentiment}, finance~\cite{mishev2020evaluation}, and healthcare~\cite{zunic2020sentiment}. In the mean time, it requires no specific domain knowledge from our human subjects. However, determining the sentiment in lengthy and unstructured text can be time-consuming and laborious for humans~\cite{zhang2020long,sheng2021efficient}, while 
AI technologies hold promise in streamlining this process by providing automated suggestions. Therefore, the AI-assisted sentiment analysis task we used in this experiment reflects the real-world scenarios where people utilize their general human intelligence to provide sentiment labels for texts, mostly by verifying the labels produced by automatic AI technologies. These scenarios can often been found in AI-assisted human labeling~\cite{desmond2021increasing,ashktorab2021ai,desmond2022ai} and human-in-the-loop machine learning pipelines~\cite{mosqueira2023human,monarch2021human,wu2022survey}.  
Similar tasks have also been used in previous studies to investigate human behavior in AI-assisted decision-making~\cite{chu2020visual,schmidt2019quantifying,hase2020evaluating},  {and to explore ways to promote humans' appropriate reliance on AI in AI-assisted decision-making}~\cite{bansal2021does}.  

{We note that for the IMDB dataset from which we draw our decision-making tasks, the ground-truth label for a movie review's sentiment is decided by the reviewer's {\em own} star rating of the movie (on a scale of 1 to 10)  associated with their review text. As described in{~\cite{maas-EtAl:2011:ACL-HLT2011}}, the star ratings are converted to a binary label by mapping a review with a rating $\le4$ out of 10 as a negative review, while a review with a rating $\ge7$ out of 10 as a positive review, and reviews that potentially have ambiguous sentiment (i.e., with ratings between 4 and 7) are not included in the dataset. In other words, the ground-truth labels of our decision making tasks are established by the reviewers of the movie, rather than through aggregating labels generated by crowd workers in a crowdsourced annotation effort\footnote{We acknowledge that in general, sentiment analysis tasks have a degree of subjectivity~\cite{gordon2022jury,davani2022dealing}. However, we believe that by asking subjects to determine the \textit{polarity} of the sentiment instead of \textit{specific emotion} contained in the review, and by using a dataset that does not include the middle-range rated, potentially ambiguous reviews, we minimize the possibility that the ground-truth label (i.e., the correct decision in a decision-making task) is subject to debate or unreliable.}. For a complete list of 20 movie reviews that we used in our experiment and their ground-truth sentiment labels,  please see the supplemental materials.

\vspace{2pt}
\noindent \textbf{The AI model used in the task.} On each sentiment analysis task, all subjects in our experiment were presented with the prediction given by the {\em same} AI model. In particular, to obtain this AI model, we fine-tuned a pre-trained RoBERTa model~\cite{liu2019roberta} from the Huggingface's transformers library---First, from the IMDB movie review dataset, we sampled a subset of 5,000 movie reviews to be used as our training set, as well as another subset of 500 reviews as the test set. We used the representation embeddings of the last layer of the pre-trained RoBERTa model as the input of a multi-layer perceptron and tuned parameters of both the pre-trained model and the perceptron based on the training data. Our final model achieved an accuracy of 77.6\% on the held-out test set. On the set of 20 movie reviews we used in our experiment, the model's accuracy was 75\% (i.e., correct on 15 tasks and incorrect on 5 tasks), which closely reflected the model's overall performance on the test set. We also intentionally did not train a model with very high performance so that we would have sufficient data to understand how subjects behave in AI-assisted decision-making both when the AI model is correct and wrong (e.g., analyze subjects' under-reliance and over-reliance separately).


\subsection{Experimental Treatments}\label{sec:design}


\ignore{
\begin{figure*}[t]
\centering

\includegraphics[width=\textwidth]{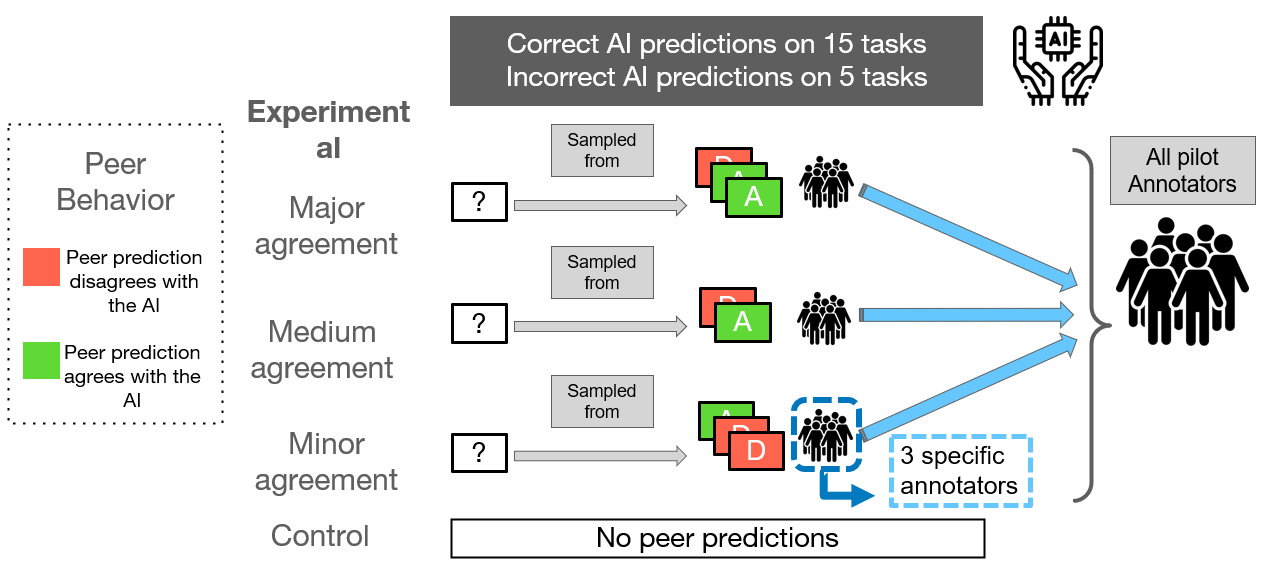}
\vspace{-15pt}
\caption{The design and treatment implementation of Experiment 1.  }~\label{fig:exp1:procedure}
\vspace{-20pt}
\end{figure*}
}

In total, we created four treatments for our experiment by varying the presence of  second opinions from peers and the level of agreement between peers' judgements and the AI model's predictions. 

Specifically, to enable the presence of second opinions 
from real human decision-makers to ensure ecological validity, we first ran a pilot study in which 34 MTurk workers were recruited to review the same set of 20 movie reviews that we selected for our experiment. These workers were asked to determine the sentiment of each review {\em independently}, i.e., without seeing our AI model's prediction. For each worker, we then computed the fraction of tasks in which their independent judgement was the same as our AI model's prediction across all 20 tasks; we denoted this fraction as the worker's ``level of agreement'' with the AI model. 
After each worker's level of agreement with the AI model was computed, we identified three subsets of workers from the entire pool of 34 workers, with each subset containing three workers---The first subset contained the three workers who agreed with the AI model most frequently, and we referred to them as the ``{\em high agreement peers}''; the second subset contained three workers who agreed with the AI model on about 50\% of the tasks, and we referred to them as the ``{\em medium agreement peers}''; finally, the last subset contained the three workers who agreed with the AI model least frequently, whom we referred to as the ``{\em low agreement peers}''\footnote{For the subset of high, medium, and low agreement peers, the average level of agreement between the crowd workers and the AI model was 76.67\%, 50\%, and 30\%, respectively.}.

Utilizing these three sets of ``peers'' that we identified from our pilot study, we designed the following 4 treatments:
\begin{itemize}
    \item \textbf{Treatment 1 (Control)}: Subjects had access to the predictions of the AI model when completing each task. However, they did not see any judgement made by other peer workers.
    \item \textbf{Treatment 2 (High agreement)}: Subjects had access to predictions made by the AI model when completing each task. In addition, on each task, we randomly selected a worker from the set of three {\em high agreement peers}, and presented the selected worker's judgement on that task to the subject as a second opinion\footnote{ {While in this experiment, the second opinions presented to subjects have already been collected from crowd workers in the pilot study, in reality, they can be solicited from peers in the real time when decision-maker is about to make their decision on a task. We decided to collect second opinions ahead of time in this study to simplify the experimental procedure and to enable the quantification of the level of agreement between second opinions and the AI model.}}.
    \item \textbf{Treatment 3 (Medium agreement)}: Subjects had access to predictions made by the AI model when completing each task. In addition, on each task, we randomly selected a worker from the set of three {\em medium agreement peers}, and presented the selected worker's judgement on that task to the subject as a second opinion.
    \item \textbf{Treatment 4 (Low agreement)}: Subjects had access to predictions made by the AI model when completing each task. In addition, on each task, we randomly selected a worker from the set of three {\em low agreement peers} and presented the selected worker's judgement on that task to the subject as a second opinion.
\end{itemize}

Figure~\ref{fig:exp1:interface} shows an example of the task interface for treatments where the second opinions from peers are presented (i.e., Treatment 2, 3, or 4). With this design, we expect that subjects in Treatment 2 will find the second opinions generated by peer workers agree with the AI model more frequently than subjects in Treatment 3, who in turn will observe a higher level of agreement between the peers and the AI model than subjects in Treatment 4. 
The manipulation of the level of agreement between the AI model's decision recommendation and the second opinion across treatments is crucial, as it reflects the degree of conflicts between two ``advisors''. 
In the traditional advice-taking literature, the 
conflicts between advisors have been found to have nuanced impacts on people's advice taking behavior, and we expect the same holds true for AI-assisted decision making settings as well. 
In particular, in AI-assisted decision making, when the human decision maker receives the second opinion solicited by the system from another {\em random} peer, it is unclear ex-ante what the level of conflict (or disagreement) between that peer and the AI model's recommendation would be.  
Thus, by creating three experimental treatments where the second opinion was solicited from peers with varying levels of agreement with AI, we are able to obtain a comprehensive understanding 
of how the level of conflict exhibited between the random peer and the AI model moderates the effects of the second opinion on the human decision maker's behavior and performance in AI-assisted decision making. 

\begin{figure*}[t]
\centering

\includegraphics[width=0.6\textwidth]{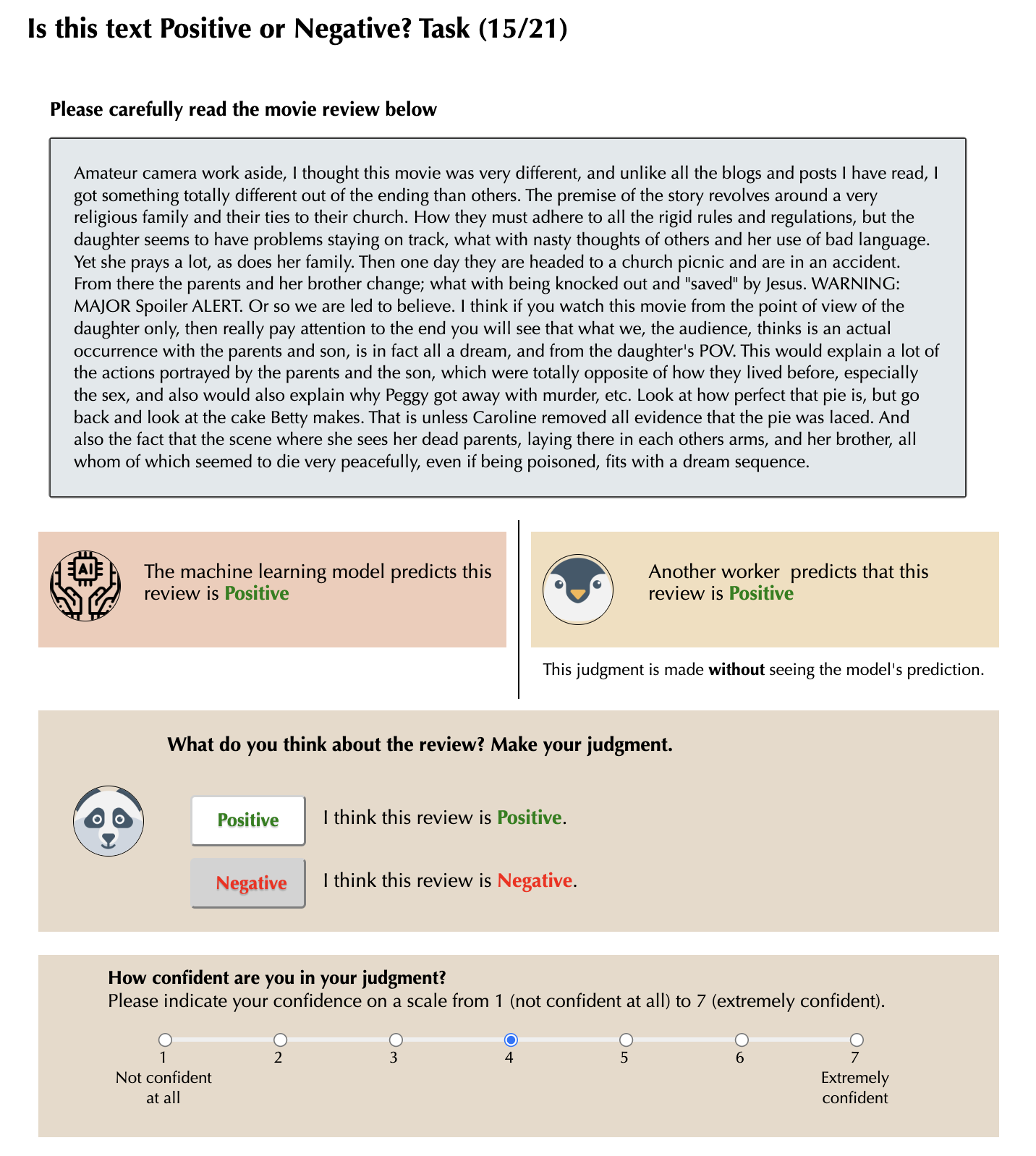}
\vspace{-10pt}
\caption{An illustration of our task interface in Experiment 1. In this example, the treatment is presenting both the AI model's prediction (positive) and the second opinion from a peer worker (negative).}
~\label{fig:exp1:interface}
\vspace{-10pt}
\end{figure*}



\subsection{Experimental Procedure}
Our experiment was posted on Amazon Mechanical Turk (MTurk) as a human intelligence task (HIT). The HIT was open to workers in the U.S. only, and each worker could only take the HIT once. Each HIT contained the same 20 movie review tasks, which were arranged in a random order. In addition, we included an attention check question in our HIT, in which the subject were instructed to select a pre-specified option. We only considered the data generated by subjects who passed the attention check as valid data in our analysis.  

Upon arrival at the HIT, subjects were randomly assigned to one of the four treatments. Subjects first received instruction on the movie review task. In order to show that they understood how to complete the movie review tasks, subjects needed to complete a qualification task, in which they were asked to review a simple movie review and determine its sentiment. Subjects could proceed to the actual experiment only if they answered the qualification question correctly. In the actual experiment, subjects were first asked to pick an avatar to represent themselves throughout the experiment. Then, as we have discussed in Section~\ref{sec:task}--\ref{sec:design}, subjects completed the 20 movie review tasks;  depending on the treatment they were assigned, on each task, they saw the decision recommendation generated by our AI model and possibly by other peer workers. 
In each task, beyond making a decision on the movie review's sentiment, subjects were also asked to indicate how confident they were in their decision using a 7-point Likert scale from 1 (``not confident at all'') to 7 (``extremely confident'').
After completing all 20 tasks, subjects needed to fill out an exit-survey. In this survey, in addition to general demographic information questions (e.g., age, gender, education), we also asked subjects to estimate the AI model's, the peers' (if applicable), and their own accuracy in analyzing movie review sentiment across the 20 tasks in the HIT. 

The base payment of our experiment is \$1.2. To encourage subjects to carefully deliberate about the decision recommendations made by the AI model and the peers, we also provided a performance-based bonus to subjects---If the subject's accuracy in our HIT was higher than 65\%, we paid them an additional 5-cent bonus for each correct prediction they made; thus, subjects could earn up to \$1 bonus in our experiment in addition to the base payment\footnote{ {The median time that subjects spent on Experiment 1 was 176 seconds, leading to a median hourly payment of \$24.5.}}.

\subsection{Analysis Methods}
\label{methods}

To understand how second opinions from human peers affect people's behavior and performance in AI-assisted decision-making, we pre-registered a set of dependent variables for this experiment\footnote{The pre-registration document can be found at: \url{https://aspredicted.org/36J_RHS}.  {All of our experiments were approved by the IRB of the authors' institution.}}. Specifically, to examine how peers' judgements change people's reliance on AI models in AI-assisted decision-making, and whether these changes are desirable or not, we consider the following dependent variables: 

\begin{itemize}
     \item  \textbf{Overall reliance}: 
     The chance for a subject's decision to be the same as the AI model's prediction.
    \item \textbf{Over-reliance}: 
    The chance for a subject's decision to be the {\em same} as the AI model's prediction, when the AI model's prediction was {\em incorrect}.
    \item \textbf{Under-reliance}:
    The chance for a subject's decision to be {\em different} from the AI model's prediction, when the AI model's prediction was {\em correct}.
    \item \textbf{Appropriate reliance}:
    The chance for a subject's decision to be the same as the AI model's correct prediction or different from the AI model's incorrect prediction;
    this effectively represents the subject's {\em decision accuracy}.
\end{itemize}

A subject's overall reliance quantifies the subject's reliance behavior in AI-assisted decision-making without differentiating whether such reliance is desirable. We then used over-reliance, under-reliance, and appropriate reliance to understand whether the reliance behavior that the subject exhibited was desirable or not. Intuitively, a desirable reliance behavior requires lower levels of over-reliance and under-reliance, and higher levels of appropriate reliance\footnote{We acknowledge that recent studies have introduced more sophisticated metrics for measuring appropriate reliance on AI, such as ``
relative self-reliance'' (RSR) and ``relative AI reliance'' (RAIR)~\cite{schemmer2022should,schemmer2023appropriate}. 
However, we did not use these metrics in our study for several reasons. First, our study concerns a AI-assisted decision making setting commonly used in the real life (especially in AI-assisted labeling pipelines) where decision makers are presented with the AI model's decision recommendation upfront, without having to register their independent decisions first. However, the computation of RSR and RAIR requires the knowledge of these independent human decisions. Second, RSR and RAIR focus on quantifying the appropriateness of reliance for only those cases where humans' independent decision disagrees with the AI recommendation. One may argue that whether humans accept the AI recommendation when it agrees with their own independent judgement also carries important information (e.g., if humans decide to switch away from this agreement, it may imply a very high level of distrust to AI). RSR and RAIR cannot capture this information, but the metrics we used can. 
}.



In addition, to understand how second opinions from peers affect people's performance in AI-assisted decision-making beyond their decision accuracy (i.e., appropriate reliance), we included a few more dependent variables related to subjects' decision time and decision confidence:

\begin{itemize}
    \item \textbf{Decision time}: The amount of time that a subject spent on a task\footnote{As per our pre-registration, decision times that were longer than the 95\% percentile of the decision time distribution obtained from all subjects on all tasks were treated as outliers and removed from the analysis; for example, in Experiment 1, 419 decision time records were excluded from a total of 8,560 records. 
    }. 
    \item \textbf{Confidence in correct decisions}: The average level of confidence subjects reported in a task if the subject's decision on that task was correct; this was computed for each of the 20 tasks.
    \item \textbf{Confidence in incorrect decisions}: The average level of confidence subjects reported in a task if the subject's decision on that task was incorrect; this was computed for each of the 20 tasks.
\end{itemize}

Holding everything else equal, we may consider a subject's performance to be better if they spend less time on the task, and become more confident in their correct decisions while less confident in their incorrect decisions.  

Based on our pre-registration, for dependent variables related to reliance and decision time, we conducted the one-way analysis of variance (ANOVA) to examine whether there are any significant differences in them across the 4 experimental treatments. When a significant difference was found, we used the Tukey HSD tests to conduct post-hoc pairwise comparisons. 
For dependent variables related to decision confidence, since they were aggregated on each of the 20 tasks\footnote{Since subjects in different treatments received different second opinions (if any), they might be correct/incorrect on different sets of tasks. This means that directly comparing subjects' average decision confidence in their correct (or incorrect) decisions across treatments without aggregating to the task level can be misleading, because the comparison may occur between decision confidence reported for different distributions of tasks.}, we used repeated measures ANOVA to examine whether a significant difference exists across experimental treatments, and pairwise paired t-tests with Bonferroni corrections were used as our post-hoc analysis to identify pairs of treatments that exhibit significant differences.


\subsection{Experimental Results}
In total, 428 subjects took our experiment HIT and passed the attention check (56.8\% self-identified as male, 41.1\% self-identified as female, and the most frequent age group reported by subjects was 25-34). To begin with, for the three treatments with peer judgements (i.e., Treatments 2--4), we checked the level of agreement between the AI model and the actual peer judgements presented to subjects. 
The average fraction of tasks in which the peer worker's judgement agreed with the AI model was 0.77, 0.51, 0.30 for treatments with high, medium, and low agreement peers, respectively, and a one-way ANOVA test confirms that the level of agreement between peers and the AI model across these three treatments is significantly different ($F(3,424) = 790.34, p < 0.001$). This indicates that we successfully varied the peer-AI agreement level through our experimental design.

\subsubsection{Effects on Reliance on AI} First, we look into how the presence of second opinions from human peers 
affects decision-makers' reliance on the AI model in AI-assisted decision-making. 

\vspace{2pt}
\noindent\textbf{\em Second opinions from peers decrease people's overall reliance on the AI model}. Figure~\ref{fig:exp1:reliance} shows subjects' average level of overall reliance on the AI model across the four treatments. Visually, it is clear that the presence of second opinions from human peers results in a {\em decrease} in people's overall reliance on AI. Also, the more the peers disagree with the AI model, the more the reliance decreases. A one-way ANOVA test confirms that the difference in subjects' overall reliance across different treatments is statistically significant ($F(3, 8556) = 28.31, p < 0.001$). The post-hoc Tukey HSD test suggests that subjects in all treatments with peer judgements are less likely to rely on the AI model than those in the control treatment (i.e., control vs. high agreement: $p < 0.001$, Cohen's $d = 0.17$; control vs. medium agreement: $p < 0.001$, Cohen's $d = 0.21$; control vs. low agreement: $p < 0.001$, Cohen's $d = 0.28$). In addition, the overall reliance difference shown between the two treatments with high agreement peers and low agreement treatment peers is also found to be significant (high agreement vs. low agreement: $p < 0.001$, Cohen's $d = 0.11$).

\vspace{2pt}
\noindent \textbf{\em Second opinions from human peers 
lead to lower over-reliance, higher under-reliance, and do not significantly affect the appropriate reliance}. 
To understand whether the decrease in people's overall reliance on the AI model brought up by second opinions from peers is desirable, we further examine people's over-reliance, under-reliance, and appropriate reliance on AI separately. 
First, Figure~\ref{fig:exp1:overreliance}
compares subjects' over-reliance on the AI model across the four treatments.  { It suggests that the presence of second opinions from peers helps subjects \textit{reduce} their over-reliance on the AI model, especially when the peers' judgements have a relatively low level of agreement with the AI.} 
The one-way ANOVA test indicates that the differences in subjects' over-reliance are significant across treatments ($F(3, 2136) = 17.27, p < 0.001$).  {Post-hoc Tukey HSD tests further show that these significant differences exist between the control treatment and every experimental treatment with peer judgements (control vs. high agreement: $p < 0.001$, Cohen's $d = 0.26$; control vs. medium agreement: $p < 0.001$, Cohen's $d = 0.31$; control vs. low agreement: $p < 0.001$, Cohen's $d = 0.44$). A significant difference is also found between the treatment with high agreement peers and the one with low agreement peers ($p = 0.010$, Cohen's $d = 0.18$).  }

While second opinions from peers bring about the benefit of decreased levels of over-reliance, 
these benefits also come with a cost. Specifically, Figure~\ref{fig:exp1:underreliance} shows subjects' average levels of under-reliance on the AI model in the four treatments. 
Here, we also find a significant difference across treatments (one-way ANOVA: $F(3, 6446) = 13.84, p < 0.001$). Post-hoc Tukey HSD tests show that compared to when the second opinions are absent, subjects significantly {\em increase} their under-reliance on the AI model when they receive the peers' judgements as the second opinions, regardless of how frequently the peers' judgements agree with the AI   
(i.e., control vs. high agreement: $p < 0.001$, Cohen's $d = 0.14$; control vs. medium agreement: $p < 0.001$, Cohen's $d = 0.17$; control vs. low agreement: $p < 0.001$, Cohen's $d = 0.22$). 

Together, our results suggest that when peer-generated second opinions are presented in AI-assisted decision-making, people decrease their reliance on the AI model {\em regardless of} the AI model's prediction correctness.  
As such, when examining subjects' appropriate reliance on the AI model across different treatments (Figure~\ref{fig:exp1:approp_reliance}), we did not find that there is any significant difference.

\begin{figure*}[t]
     \centering
     \begin{subfigure}[b]{0.235\textwidth}
         \centering
     \includegraphics[width=\textwidth]{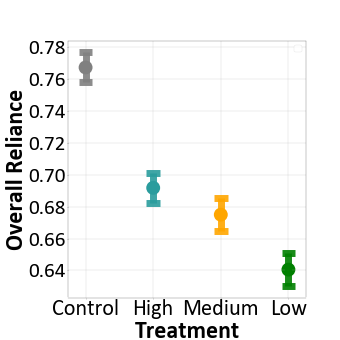}
         \caption{Overall reliance}
         \label{fig:exp1:reliance}
     \end{subfigure}
    \begin{subfigure}[b]{0.235\textwidth}
         \centering
     \includegraphics[width=\textwidth]{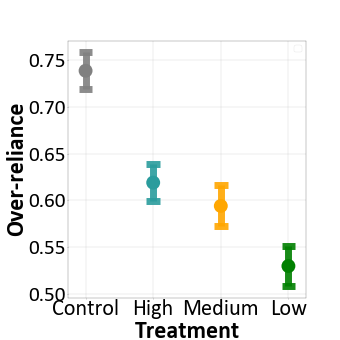}
         \caption{Over-reliance}
     \label{fig:exp1:overreliance}
     \end{subfigure}
          \begin{subfigure}[b]{0.235\textwidth}
         \centering
     \includegraphics[width=\textwidth]{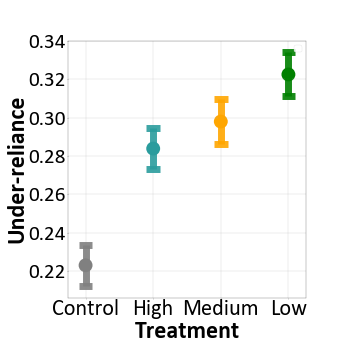}
         \caption{Under-reliance}
         \label{fig:exp1:underreliance}
     \end{subfigure}
     \centering
         \begin{subfigure}[b]{0.235\textwidth}
         \centering
     \includegraphics[width=\textwidth]{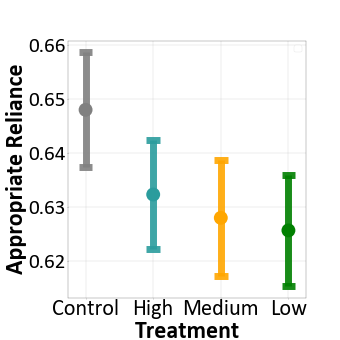}
         \caption{Appropriate reliance}
     \label{fig:exp1:approp_reliance}
     \end{subfigure}
     \vspace{-5pt}
    \caption{The effects of second opinions from human peers on subjects' overall reliance (\ref{fig:exp1:reliance}), over-reliance (\ref{fig:exp1:overreliance}),  under-reliance (\ref{fig:exp1:underreliance}), and appropriate reliance (\ref{fig:exp1:approp_reliance}) on the AI model across treatments. Error bars represent the standard errors of the mean. 
    }
        \label{fig:exp1:reliance_all}
\vspace{-12pt}
\end{figure*}

\subsubsection{Effects on decision time} 
Figure~\ref{fig:exp1:time} illustrates the average decision time subjects spent on a task. 
As expected, the presence of a second opinion 
makes subjects spend {\em more} time to make their decisions, compared to subjects in the control treatment. 
Our one-way ANOVA test result suggests that the difference in decision times across treatments is significant ($F(3, 8122) = 12.06, p < 0.001$). Pair-wise comparisons indicate that subjects who received second opinions from medium and low agreement peers spent significantly more time on a task than both those subjects who did not receive second opinions (control vs. medium agreement: $p < 0.001$, Cohen's $d = 0.17$; control vs. low agreement: $p < 0.001$, Cohen's $d = 0.15$), and those subjects who received second opinions from the high agreement peers (high agreement vs. medium agreement: $p = 0.004$, Cohen's $d = 0.10$; high agreement vs. low agreement: $p = 0.022$, Cohen's $d = 0.09$).

\subsubsection{Effects on confidence}
In Figures \ref{fig:exp1:confidence_correct} and  \ref{fig:exp1:confidence_incorrect}, we plot the average values of subjects' confidence in their correct decisions and incorrect decisions, respectively. Using repeated measures ANOVA, we detect significant differences across treatments in  subjects' confidence for both their correct decisions ($F(3,16)=6.86, p<0.001$) and their incorrect decisions ($F(3,16)=3.33, p=0.045$). Through the post-hoc pairwise t-tests with Bonferroni corrections, we find that for subjects' correct decisions, subjects in the treatment with low agreement peers had significantly higher confidence than subjects in both the control treatment ($p=0.003$, Cohen's $d=1.08$) and the treatment with high agreement peers ($p=0.002$, Cohen's $d=1.64$). 
In contrast, for subjects' incorrect decisions, none of the pair-wise comparisons are significant at the level of $p=0.05$.

\begin{figure*}[t]
 \centering
    \begin{subfigure}[b]{0.235\textwidth}
         \centering
     \includegraphics[width=\textwidth]{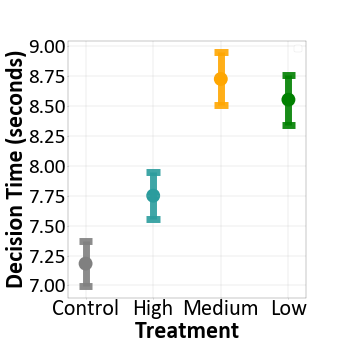}
         \caption{Decision time}
     \label{fig:exp1:time}
     \end{subfigure}
     \centering
     \begin{subfigure}[b]{0.235\textwidth}
         \centering
     \includegraphics[width=\textwidth]{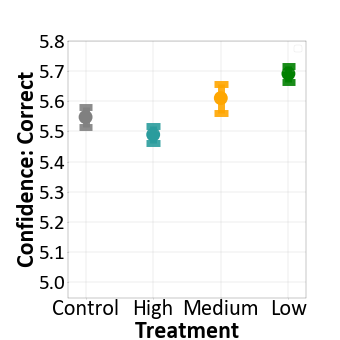}
         \caption{Confidence (Correct)}
         \label{fig:exp1:confidence_correct}
     \end{subfigure}
     \begin{subfigure}[b]{0.235\textwidth}
         \centering
     \includegraphics[width=\textwidth]{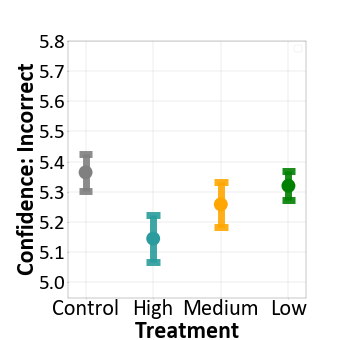}
         \caption{Confidence (Incorrect)}
         \label{fig:exp1:confidence_incorrect}
     \end{subfigure}
   
     \vspace{-5pt}
    \caption{The effects of second opinions from human peers on subjects' decision time (\ref{fig:exp1:time}), confidence in their correct decisions (\ref{fig:exp1:confidence_correct}), and confidence in their incorrect decisions (\ref{fig:exp1:confidence_incorrect}) across treatments. Error bars represent the standard errors of the mean.
    }
        \label{fig:exp1:confidence_time}
\vspace{-10pt}
\end{figure*}

\ignore{

\begin{table}[]
\centering
\begin{tabular}{|c|c|c|}
\hline
\multirow{2}{*}{} & Reliance: AI Incorrect & Reliance: AI Correct \\ \cline{2-3} 
                & Model 1    & Model 2 \\ \hline
high agreement peer \textbf{agrees} with AI ($\beta_1$)    & -0.45(*)  & -0.31        \\ \hline
medium agreement peer \textbf{agrees} with AI ($\beta_2$)   & -0.64(*) & -0.32       \\ \hline
low agreement peer \textbf{agrees} with AI ($\beta_3$)      & -0.54  & -0.46       \\ \hline
high agreement peer \textbf{disagrees} with AI ($\beta_4$)  & -1.13(***) & -1.21(***)  \\ \hline
medium agreement peer \textbf{disagrees} with AI ($\beta_5$)& -0.86(**) & -1.06(***)  \\ \hline
low agreement peer \textbf{disagrees} with AI ($\beta_6$)   & -1.23(***) & -1.05(**)  \\ \hline

Peer Accuracy  ($\beta_7$)   & 0.49 & -0.42  \\ \hline
\end{tabular}
\caption{Regression coefficients of independent variables indicating the which treatment (high, medium, low) a prediction is made in and whether the AI model agreed with the peer prediction or not (agree, disagree) on reliance. Data is split into 2 subgroups based on whether the AI model is correct. The regression models take each task and subject as a random effect and the accuracy of the peer that subjects experienced as a covariate.. p-value significance: (.) p < 0.1, (*) p < 0.05, (**) p < 0.01, (***) p < 0.001}
\label{table:regression:reliance_time}

\end{table}
}

\subsubsection{Exploratory Analyses} 
Finally, we conduct a set of exploratory analyses to better understand the reasons behind our findings, especially on why the presence of second opinions generated by human peers results in both decreased over-reliance and increased under-reliance on AI. 
Detailed analyses can be found in the supplemental materials. 


Here, we highlight a set of analysis in which we attempt to 
understand how people's reliance on the AI model is affected by the comparison between the AI recommendation and the peer-generated second opinion {\em at the level of individual tasks}. Specifically, when a second opinion from a peer is presented on a particular task, the peer may agree or disagree with the AI model. How does this agreement or disagreement 
affect people's reliance on the AI model on that task? 
And how does this effect vary with the peer's overall level of agreement with the AI model across many tasks and the AI model's correctness on that task? Answering these questions can provide more nuanced insights into how people react to peer-generated second opinions on the task level, and may help explain some of our observations. 

Therefore, we construct mixed-effect regression models to predict whether a subject's decision would be the same as the AI model's prediction on a task (i.e., whether the subject would rely on the AI model on a task). In these regression models, whether the peer---who may have a high, medium, or low level of agreement with the AI model overall---agrees or disagrees with the AI model on the current task is used as the fixed effect, while the subject and the decision-making task are treated as the random effects\footnote{As whether the subject relies on the AI model or not on a task is binary, we build mixed-effect logistic regression models.}. 
We fit the regression models separately for tasks on which the AI model is correct and incorrect, and results are reported in Table~\ref{table:regression:reliance_time}.
Inspecting the estimated coefficients for those independent variables that indicate the peer {\em agrees} with the AI model on a task (i.e., $\beta_1$--$\beta_3$), we find that, surprisingly, the agreement between peers and the AI model on a task does {\em not} nudge people into relying on the AI model more (i.e., none of the estimated $\beta_1$--$\beta_3$ are significantly positive). On the other hand, we also find that in both models, all the estimated coefficients for those independent variables that indicate the peer {\em disagrees} with the AI model on a task (i.e., $\beta_4$--$\beta_6$) are significantly negative. This means that once observing that the peer disagrees with the AI model on a task, people significantly decrease their reliance on the AI model regardless of the AI model's correctness on that task. 

\begin{table}[t]
\centering
\resizebox{\linewidth}{!}{

\begin{tabular}{c|c|c}
\toprule
\multirow{2}{*}{} & \textbf{Reliance: AI Correct} & \textbf{Reliance: AI Incorrect}  \\ 
                & \textbf{(Model 1)}    & \textbf{(Model 2)}
\\\midrule
Intercept ($\beta_0$) &  1.93*** &   1.34$^{***}$        \\ \hline

high agreement peer \textbf{agrees} with AI ($\beta_1$)      & -0.31  & -0.46$^{*}$      \\ \hline
medium agreement peer \textbf{agrees} with AI ($\beta_2$)    & -0.25   & -0.75**    \\ \hline
low agreement peer \textbf{agrees} with AI ($\beta_3$)   & -0.34    & -0.69**        \\ \hline
high agreement peer \textbf{disagrees} with AI ($\beta_4$) & -1.21*** & -1.14***   \\ \hline
medium agreement peer \textbf{disagrees} with AI ($\beta_5$) & -0.98*** & -0.96***   \\ \hline
low agreement peer \textbf{disagrees} with AI ($\beta_6$) & -0.93***  & -1.37***   \\ \bottomrule
\end{tabular}
}
\vspace{2pt}
\caption{
Understanding how subjects' reliance on a task is influenced by the agreement or disagreement between the AI model and the peers on that task, while the peers may have different overall frequencies to agree with the AI model.
Mixed-effect regression models are built for tasks that the AI model is correct (Model 1) or incorrect (Model 2) separately, and each task and each subject is treated as a random effect.
Coefficients estimated are reported. 
 *, **, *** indicate significance levels of $0.05, 0.01,$ and $0.001$, respectively.}
\label{table:regression:reliance_time}
\vspace{-10pt}
\end{table}

Together, these results present a more detailed characterization of how the presence of second opinions from human peers affects people's reliance on AI models---When the AI model is incorrect on a task, the presence of second opinions significantly reduces people's reliance on the AI {\em no matter} whether the second opinions align with the AI model's recommendation or not, leading to lower levels of {\em over-reliance}. When the AI model is correct on a task, however, the presence of (incorrect) second opinions that disagree with the AI also significantly reduces people's reliance on the AI, resulting in higher levels of {\em under-reliance}. Finally, as the decreases in people's reliance on the AI model caused by peer-AI disagreements are larger than those caused by peer-AI agreements ($|\beta_4|, |\beta_5|, |\beta_6|>|\beta_1|, |\beta_2|, |\beta_3|$), it is natural that second opinions from peers who have a lower level of overall agreement with the AI model bring about lower levels of over-reliance and higher levels of under-reliance. 


\section{Experiment 2: Second Opinions From Different Sources}

Our Experiment 1 shows that providing second opinions from {\em human peers} to decision-makers in AI-assisted decision-making have significant impacts on decision-makers' reliance behavior and some aspects of their decision-making performance. Naturally, one may wonder to what extent these effects are specific to second opinions generated by human peers. For example, if the second opinions are claimed to be produced by another AI model, would we see a similar or different effect?

To answer this question, we conducted our second pre-registered, randomized human subject experiment\footnote{The pre-registration can be found at: \url{https://aspredicted.org/X9H_CM6}.}, where subjects were again asked to complete the same set of 20 sentiment analysis tasks, and with the assistance from the same AI model, as those used in Experiment 1. 


\subsection{Experimental Design}
\subsubsection{Experimental Treatments and Procedure} Utilizing the same set of peer judgements as those collected in Experiment 1, we created 5 treatments for Experiment 2. In the control treatment, subjects completed AI-assisted sentiment analysis tasks without the access to any second opinions. On the other hand, second opinions were provided to subjects in the other four experimental treatments, which were arranged in a 2 by 2 factorial design varying along the following two factors:

\begin{itemize}
    \item \textbf{The level of agreement between the AI model and the second opinion}: 
    On each task, the second opinion presented to the subject was randomly sampled from the pool of judgements made by {\em high agreement peers} or {\em low agreement peers}.
    \item \textbf{The stated source of the second opinion}: Subjects were told that the second opinion was generated by {\em another crowd worker}, or {\em another AI model} that was trained using a different algorithm than the primary AI model that provides the decision recommendation.
\end{itemize}

The procedure of this experiment was identical to that of Experiment 1, while subjects who had participated in Experiment 1 was not allowed to take this experiment. 



\subsubsection{Analysis Methods}
We used the same dependent variables as those used in Experiment 1 (see Section \ref{methods} for details). Following the standard practice to analyze experimental data where the control treatment does not fit into the factorial design~\cite{himmelfarb1975you}, for each dependent variable, we first conducted a one-way ANOVA test to examine whether significant differences exist across all treatments. If so, we then performed the 
post-hoc Tukey HSD tests 
to compare each experimental treatment with the control treatment. 
We then focused on the four experimental treatments to understand how the agreement level between second opinions and the AI model, as well as the stated source of second opinions, affect the dependent variables. We did so by conducting two-way ANOVA tests. 

\subsection{Experiment Results}

In total, 516 subjects participated in Experiment 2 and passed the attention check (65.3\% self-identified as male, 30.8\% self-identified as female, and the most frequent age group reported by subjects was 25-34)\footnote{The median time subjects spent in Experiment 2 was 206 seconds, and the median hourly payment is \$20.97. }. Again, as a sanity check, 
we confirmed that the second opinions presented to subjects in the high agreement treatments agreed with the AI model significantly more than those presented to subjects in the low agreement treatments ($p < 0.001$).
For brevity, in the rest of this section, we focused on reporting the results on decision-makers' reliance behavior, and results on decision-makers' decision time and confidence are included in the supplementary material. 

\subsubsection{Effects on subject's reliance on AI}

We start by analyzing how subjects' reliance on the AI model's decision recommendation differs across all treatments, and then analyze the main effects of the two factors, i.e., the agreement level between second opinions and the AI model, and the stated source of second opinions. 

\vspace{2pt}
\noindent \textbf{\em Second opinions from both sources lead to decreased overall reliance and over-reliance, increased under-reliance, and sometimes decreased appropriate reliance on AI.} Figures \ref{fig:exp3:reliance}--\ref{fig:exp3:approp_reliance} compare subjects' overall reliance, over-reliance, under-reliance, and appropriate reliance on the AI model across the five treatments, respectively. 
One-way ANOVA test results suggest that significant differences exist across the five treatments with respect to all four aspects of reliance 
(overall reliance: $F(4, 10315) = 18.95, p < 0.001$; over-reliance: $F(4, 2575) = 8.19, p < 0.001$; under reliance: $F(4, 7735) = 8.19, p < 0.001$; appropriate reliance: $F(4, 10135) = 3.33, p = 0.010$). 
Through post-hoc Tukey HSD tests, we find that in all four treatments with the access to second opinions, subjects' overall reliance on AI is significantly lower than that in the control treatment ($p<0.001$), while subjects' under-reliance is significantly higher than that in the control treatment ($p<0.001$). For over-reliance, except for those in the ``high agreement--AI source'' treatment (i.e., second opinions are claimed to come from another AI model and have a high level of agreement with the primary AI model's decision recommendations), 
subjects in all other treatments with the access to second opinions show significantly lower levels of over-reliance than those subjects in the control treatment ($p<0.05$). Finally, subjects in the ``high agreement--AI source'' and ``low agreement--human source'' treatments are found to have a significantly lower level of appropriate reliance than subjects in the control treatment ($p<0.05$).  



\begin{figure*}[t]
     \centering
     \begin{subfigure}[b]{0.235\textwidth}
         \centering
     \includegraphics[width=\textwidth]{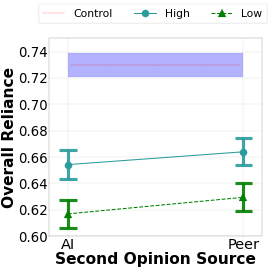}
         \caption{Overall reliance}
         \label{fig:exp3:reliance}
     \end{subfigure}
    \begin{subfigure}[b]{0.235\textwidth}
         \centering
     \includegraphics[width=\textwidth]{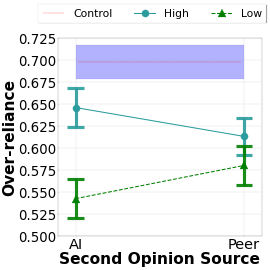}
         \caption{Over-reliance}
     \label{fig:exp3:overreliance}
     \end{subfigure}
          \begin{subfigure}[b]{0.235\textwidth}
         \centering
     \includegraphics[width=\textwidth]{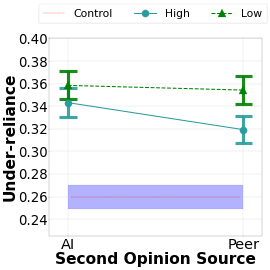}
         \caption{Under-reliance}
         \label{fig:exp3:underreliance}
     \end{subfigure}
     \centering
         \begin{subfigure}[b]{0.235\textwidth}
         \centering
     \includegraphics[width=\textwidth]{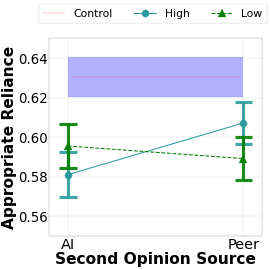}
         \caption{Appropriate reliance}
     \label{fig:exp3:approp_reliance}
     \end{subfigure}
     \vspace{-5pt}
    \caption{The effects of second opinions from different sources (i.e., human peers or another AI model) and with different levels of agreement with the primary AI model on subjects' overall reliance (\ref{fig:exp3:reliance}), over-reliance (\ref{fig:exp3:overreliance}),  under-reliance (\ref{fig:exp3:underreliance}), and appropriate reliance (\ref{fig:exp3:approp_reliance}) on AI. Error bars and error shades represent the standard errors of the mean. 
    }
        \label{fig:exp3:reliance_all}
            \vspace{-10pt}

\end{figure*}

\vspace{2pt}
\noindent \textbf{\em The agreement level between second opinions and the AI model significantly affects the effects of second opinions, while the source of second opinions does not.
} 
Our two-way ANOVA tests reveal that the level of agreement between second opinions and the AI model has significant main effects on subjects' overall reliance, over-reliance, and under-reliance on the AI recommendation, but not the appropriate reliance. Specifically, when the second opinions agree with AI recommendation less frequently, subjects showed a significant decrease in their overall reliance on AI ($p<0.001$) and over-reliance on AI ($p=0.002$), while they exhibited a significant increase in their under-reliance ($p=0.038$). On the other hand, we do not find significant main effect of the source of second opinions on any of these four reliance metrics, and we do not detect any significant interactions between the two factors either. In other words, the source of the second opinions does not appear to play a major role in influencing how decision-makers would utilize the second opinions in AI-assisted decision making.

 \subsubsection{Exploratory Analysis} 
 Similar as that in Experiment 1, we now zoom in to the level of individual tasks to 
 understand how decision-makers' reliance on the  AI model on a task is affected by (1) the agreement between the second opinion and the AI recommendation on that task, (2) the overall level of agreement between the second opinions and the AI recommendations across many tasks, and (3) the stated source of second opinions. 

 Again, we construct mixed-effect regression models to predict whether a subject's decision would be the same as the AI recommendation on a task. Compared to the exploratory analysis in Experiment 1, here, we conjecture that each fixed effect may differ based on the stated source of the second opinions. Thus, we create additional fixed effect terms to capture the effects on reliance that are caused {\em exclusively} by the fact that the second opinions are claimed to be produced by human peers ($\beta_5$--$\beta_8$).   
 The regression results are reported in Table \ref{table:exp2:regression:reliance_time}. 
 First, we note that the estimated coefficients of $\beta_1$--$\beta_4$
 are all negative in both models.
 This is consistent with what we have observed previously in Experiment 1, which again suggests that one reason that may keep decision-makers from utilizing second opinions to improve their appropriate reliance on AI (i.e., decision accuracy) could be the presence of second opinions on tasks where the AI recommendation is {\em correct}. Moreover, we also notice that the estimated coefficients of $\beta_5$--$\beta_8$ are almost always insignificant (except for $\beta_6$ in Model 2). This means that the fact that second opinions are produced by human peers brings about very limited additional effects on subjects' reliance on AI recommendations, again implying that the impacts of second opinions do not vary much with their stated source. 

\begin{table}[t]
\centering
\resizebox{\linewidth}{!}{
\begin{tabular}{c|c|c}
\toprule
\multirow{2}{*}{} & \textbf{Reliance: AI Correct} & \textbf{Reliance: AI Incorrect}  \\ 
                & \textbf{(Model 1)}    & \textbf{(Model 2)}
\\\midrule
Intercept ($\beta_0$) &  1.75$^{***}$ &   1.11$^{***}$        \\ \hline
high agreement second opinion \textbf{agrees} with AI ($\beta_1$)      & -0.47$^{*}$  & -0.32$^{*}$      \\ \hline
low agreement second opinion \textbf{agrees} with AI ($\beta_2$)      & -1.53$^{***}$  & -0.34      \\ \hline
high agreement second opinion \textbf{disagrees} with AI ($\beta_3$)      & -0.53$^{*} $ & -0.46
\\ \hline
low agreement second opinion \textbf{disagrees} with AI ($\beta_4$)      & -0.92$^{***}$  & -1.08$^{***}$      \\ \hline
high agreement peer-generated second opinion \textbf{agrees} with AI ($\beta_5$)      & -0.08  & 0.21      \\ \hline
low agreement peer-generated second opinion \textbf{agrees} with AI ($\beta_6$)      & 0.47
& -0.74$^{**}$      \\ \hline
high agreement peer-generated second opinion \textbf{disagrees} with AI ($\beta_7$)      & 0.11  & 0.04      \\ \hline
low agreement peer-generated second opinion \textbf{disagrees} with AI ($\beta_8$)      & 0.06  & 0.28      \\ 
\bottomrule
\end{tabular}
}
\vspace{2pt}
\caption{Understanding how subjects' reliance on a task is influenced by the agreement or disagreement between the AI model and the second opinion on that task and whether the second opinions were claimed to be produced by human peers, while the second opinions may have different overall frequencies to agree with the AI recommendations.
Mixed-effect regression models are built for tasks that the AI model is correct (Model 1) or incorrect (Model 2) separately, and each task and each subject is treated as a random effect.
Coefficients estimated are reported. 
 *, **, *** indicate significance levels of $0.05, 0.01,$ and $0.001$, respectively. 
 }
\label{table:exp2:regression:reliance_time}
\vspace{-10pt}
\end{table}

\section{Experiment 3: 
Second Opinions Presented Only Upon Request}

Our exploratory analyses in both Experiments 1 and 2 indicate that 
the presence of second opinions---especially the disagreeing ones---on those tasks where the AI model is correct may have limited the potential of second opinions in promoting people's appropriate reliance on AI 
in AI-assisted decision making. A natural idea to overcome this limitation is 
to {\em not} present the second opinions when the AI model is correct. However, realizing this idea requires the a priori knowledge of AI correctness on each decision-making task, which is unrealistic in the real world. 

Nevertheless, in the previous two experiments, we observe some indications that people {\em may} have {\em some} capabilities to tell apart when the AI is correct and when it is wrong. 
For example, in general, people rely on the AI model more when it is correct than when it is incorrect (e.g., this can be inferred from the comparison between Figures~\ref{fig:exp1:overreliance} and~\ref{fig:exp1:underreliance}, and between Figures~\ref{fig:exp3:overreliance} and~\ref{fig:exp3:underreliance})\footnote{The chance for people to rely on AI when it is incorrect is equivalent to over-reliance, while the chance for people to rely on AI when it is correct is equivalent to 1 minus under-reliance.}.  
In light of this, can we utilize people's own perceptions of AI correctness to decide when a second opinion should be presented? For example, instead of always providing second opinions on all tasks, if these second opinions are presented only when the decision-makers actively {\em request} for them, can their presence help decrease over-reliance on AI models without increasing under-reliance?


To answer this question, we conducted our third pre-registered\footnote{The pre-registration document can be found at: \url{https://aspredicted.org/MWC_PH1}. 
}, randomized human-subject experiment.

\subsection{Experimental Design}


\subsubsection{Experimental Treatments} 

We kept the control, high agreement, and low agreement treatment as those used in Experiment 1, with only one change---for subjects in the high/low agreement treatment, instead of presenting second opinions on every task, subjects would only be presented with the second opinion on a task if they actively clicked on the ``Request'' button on it. As we did not find the source of the second opinions significantly change the effects of second opinions, in this experiment, we again told subjects the second opinions are generated by a peer crowd worker.  

\begin{figure*}[t]
     \centering
     \begin{subfigure}[b]{0.455\textwidth}
         \centering
     \includegraphics[width=\textwidth]{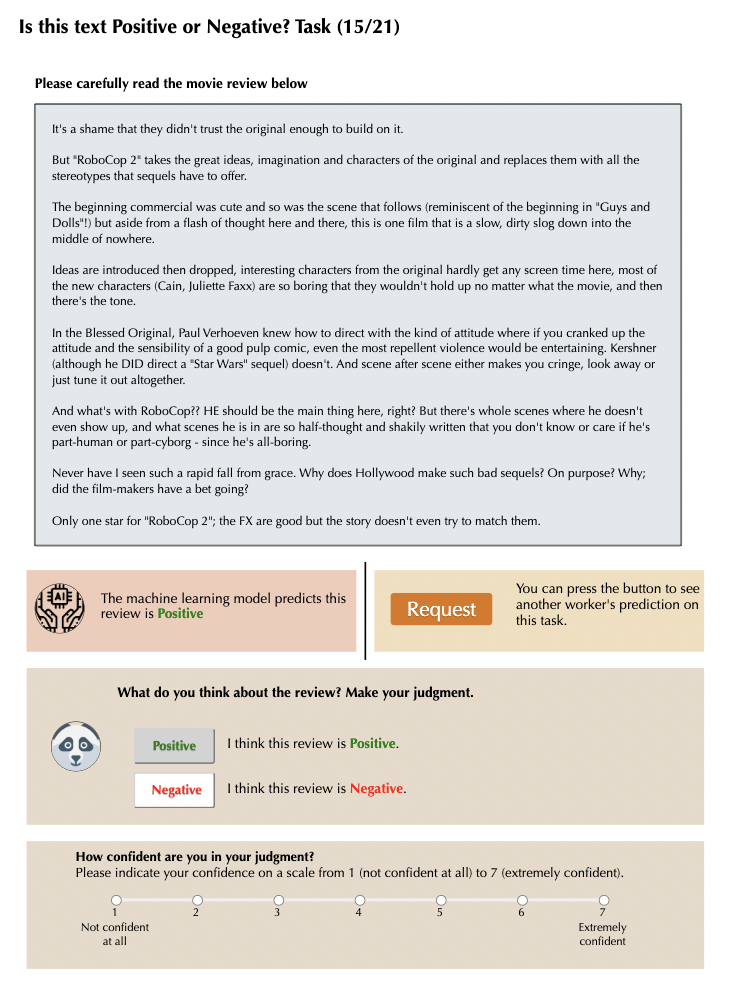}
     \vspace{-10pt}
         \caption{Before request}
     \label{fig:exp2:interface_before}
     \end{subfigure}
     \begin{subfigure}[b]{0.445\textwidth}
         \centering
     \includegraphics[width=\textwidth]{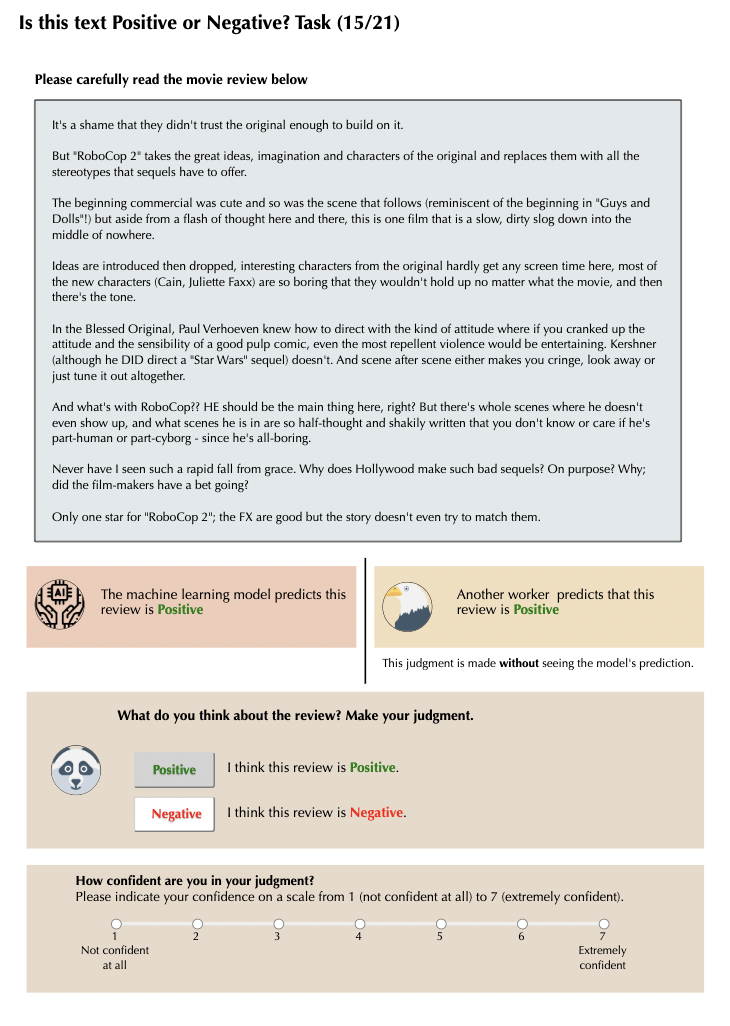}
     \vspace{-10pt}
     \caption{After request}
     \label{fig:exp2:interface_after}
     \end{subfigure}
     \vspace{-5pt}
    \caption{An example of our task interface in Experiment 3 (for treatments where subjects can solicit the second opinions), before (\ref{fig:exp2:interface_before}) and after (\ref{fig:exp2:interface_after}) subjects clicked the "Request" button. }  
        \label{fig:exp2:interface}
     \vspace{-10pt}
\end{figure*}

Figure~\ref{fig:exp2:interface} shows an example of the task interface of Experiment 3 for treatments where subjects could solicit second opinions.  


\subsubsection{Experimental Procedure} The procedure of Experiment 3 was identical to the previous experiments except for the following differences: (1) Workers who participated in previous experiments were not allowed to attend this experiment; (2) To measure subjects' tendency to engage in deliberative thinking, we added a cognitive reflection test (CRT)~\cite{frederick2005cognitive,toplak2011cognitive} in the exit-survey, which contained three mathematical questions that require people to utilize their cognitive reflection to override the intuitive, wrong answers (e.g., ``If it takes 5 machines 5 minutes to make 5 widgets, how long would it take 100 machines to make 100 widgets?''). 

\subsubsection{Analysis Methods} The dependent variables and statistical analysis methods used in Experiment 3 are the same as those outlined in Section~\ref{methods} for Experiment 1.

\subsection{Experiment Results}

In total, 336 subjects participated in Experiment 3 and passed the attention check (52.4\% self-identified as male, 45.2\% self-identified as female, and the most frequent age group reported by subjects was 25-34)\footnote{The median time subjects spent in Experiment 3 was 186 seconds, and the median hourly payment is \$23.2.}. 
Again, as a check of the effectiveness of our experimental manipulation, 
we confirmed that the actual second opinions presented to subjects upon request in the high agreement treatment agreed with the AI model significantly more than those second opinions presented to subjects in the low agreement treatment ($p < 0.001$).

\subsubsection{Effects on subjects' behavior and performance.} 
First, we conduct the main analyses on the experimental data collected from {\em all} subjects to understand that when people have the option to solicit second opinions, how their behavior (e.g., reliance on AI) and performance 
in AI-assisted decision-making change. Again, in the main paper, we focus on the decision accuracy (i.e., decision-makers' appropriate reliance on AI) as the primary performance metric; detailed analysis on other performance metrics like decision time and confidence can be found in the supplemental materials. 



\vspace{2pt}
\noindent \textbf{\em Having the option to solicit second opinions still decreases people's overall reliance and over-reliance on the AI model, and increases people's under-reliance on the AI model.}
Figures~\ref{fig:exp2:reliance}--\ref{fig:exp2:approp_reliance} compare subjects' overall reliance, over-reliance, under-reliance, and appropriate reliance on the AI model across the three treatments, respectively. It appears from the figures that the option of soliciting second opinions 
still makes people reduce their overall tendency to rely on the AI model; this  seems to decrease people's over-reliance on the AI model, increase their under-reliance, and result in a limited difference in appropriate reliance (i.e., decision accuracy), regardless of the level of agreement between the second opinions and the AI model. 
One-way ANOVA tests further show that the differences across treatments in subjects' overall reliance and under-reliance on the AI model are significant (overall reliance: $F(2, 6717) = 14.25, p < 0.001$; under-reliance: $F(2, 5037)=11.91, p<0.001$). Meanwhile, the difference across treatments in subjects' over-reliance and appropriate reliance on the AI model 
is not statistically significant at the level of $p=0.05$. Post-hoc Tukey HSD tests confirm that compared to subjects in the control treatment, those subjects who could request second opinions 
relied on the AI model significantly less in general (control vs. high agreement, $p < 0.001$, Cohen's $d = 0.12$; control vs. low agreement, $p < 0.001$, Cohen's $d = 0.15$), and they suffered from a significantly higher level of under-reliance (control vs. high agreement: $p = 0.018$, Cohen's $d = 0.12$; control vs. low agreement: $p < 0.001$, Cohen's $d = 0.16$).


\begin{figure*}[t]
     \centering
     \begin{subfigure}[b]{0.235\textwidth}
         \centering
     \includegraphics[width=\textwidth]{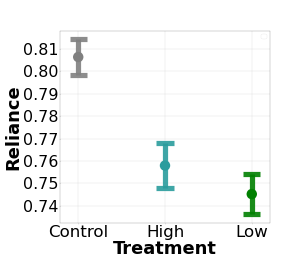}
         \caption{Overall reliance}
         \label{fig:exp2:reliance}
     \end{subfigure}
     \begin{subfigure}[b]{0.235\textwidth}
         \centering
     \includegraphics[width=\textwidth]{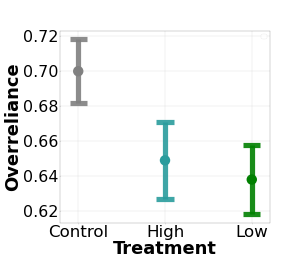}
         \caption{Over-reliance}
         \label{fig:exp2:overreliance}
     \end{subfigure}
    \begin{subfigure}[b]{0.235\textwidth}
         \centering
     \includegraphics[width=\textwidth]{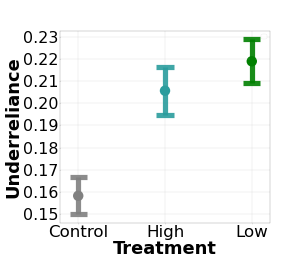}
         \caption{Under-reliance}
     \label{fig:exp2:underreliance}
     \end{subfigure}
     \centering
         \begin{subfigure}[b]{0.235\textwidth}
         \centering
     \includegraphics[width=\textwidth]{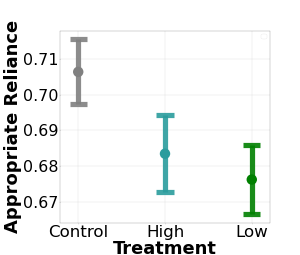}
         \caption{Appropriate reliance}
     \label{fig:exp2:approp_reliance}
     \end{subfigure}
     \vspace{-5pt}
    \caption{The effects of the optional solicitation of second opinions from peers on subjects' overall reliance (\ref{fig:exp2:reliance}), over-reliance (\ref{fig:exp2:overreliance}), under-reliance (\ref{fig:exp2:underreliance}), and appropriate reliance (\ref{fig:exp2:approp_reliance}) on the AI model across treatments. Error bars represent the standard errors of the mean.
    }
        \label{fig:exp2:reliance_all}
     \vspace{-5pt}
\end{figure*}

\subsubsection{Exploratory Analysis} 

So far, the main analyses we conduct on the experimental data collected from {\em all} subjects of Experiment 3 seem to indicate that allowing subjects to solicit second opinions still comes with the negative side effects of increasing people's under-reliance on AI models. Yet, we note that some subjects in our experiment had {\em never} requested for second opinions on any of the 20 tasks, even though they had the option to do so. So, in the following, we conduct a set of exploratory analyses to better understand subjects' behavior in requesting for second opinions. In addition, we are also interested in exploring how the presence of second opinions affects people's reliance on the AI model in AI-assisted decision making, when people actually have solicited second opinions for {\em at least once}. 

\vspace{2pt}
\noindent \textbf{\em Understanding people's behavior in requesting for second opinions}.
On average, 34.0\% of the subjects (31 subjects) in the high agreement treatment and 38.8\% of the subjects (44 subjects) in the low agreement treatment solicited second opinions for at least once among the 20 tasks. 
Taking a deeper look into where subjects solicited second opinions, we find subjects requested for second opinions slightly more when the AI model was incorrect than when the AI model was correct---For example, among subjects who solicited second opinions for at least once in the high (low) agreement treatment, the average chance for subjects to request for a second opinion on a task where the AI model was correct was 46.54\% (37.48\%), while the average chance for subjects to request for a second opinion on a task where the AI model was wrong was 50.63\% (43.11\%). 
Using a proportion test, we find that this increase in subjects' likelihood of soliciting second opinions on tasks where AI is incorrect was only marginal ($p=0.086$).  
In other words, it appears that 
the timing for people to solicit second opinions {\em may}, to some extent, reflect their perceptions of AI correctness.


Interestingly, when we split subjects in all but the control treatment into two groups based on whether they had ever requested for a second opinion from the peers, we find that the group of subjects  who requested for second opinions at least once had some different characteristics compared to the group of subjects who never requested for second opinions. For example, compared to subjects who never solicited a second opinion, subjects who solicited for second opinions at least once had lower levels of education (solicit: $M=3.88, SD=0.70$ vs. non-solicit: $M=4.10, SD=0.37$; t-test: $p=0.004$), and they also had less prior knowledge in programming (solicit: $M=2.77, SD=0.83$ vs. non-solicit: $M=3.17, SD=0.60$; t-test: $p<0.001$). 

\vspace{2pt}
\noindent \textbf{\em The effects of second opinion solicitations on people's reliance on the AI model}.
Next, we focus on only those subjects who requested second opinions for at least once, and we aim to understand how their {\em active solicitations} of second opinions changed their reliance on the AI model in AI-assisted decision-making. Given the systematic differences in demographic backgrounds between subjects who had solicited or had never solicited second opinions, directly comparing the reliance behavior of those subjects who had solicited second opinions in the high agreement or low agreement treatments with that of all subjects in the control treatment can be misleading. To ensure the robustness of our analyses, we adopt {\em matching methods} to pair up subjects with similar demographic characteristics in the control treatment and the experimental treatment, and then conduct comparisons between paired subjects.

We first conduct {\em propensity score matching}~\cite{rosenbaum1983central} for the 31 subjects in the high agreement treatment who had solicited second opinions for at least once. Specifically,
given each subject in the control treatment and the high agreement treatment, 
we characterize them using all the demographic information that they self-reported in the exit-survey (e.g., age, gender, education, prior programming knowledge, CRT score, etc.), and we build a logistic regression model to predict a subject's treatment given their features (i.e., ``covariates''). The predicted log-likelihood for a subject to belong to the high agreement treatment is thus used as the subject's ``propensity score.'' Then, for each of the 31 subjects in the high agreement treatment who requested for second opinions at least once, we identify a subject in the control treatment with the closest propensity score (with replacement) to be their ``{\em match}.'' These two subjects thus become a pair who share very similar demographic characteristics, but one subject in the pair had the chance to solicit second opinions from high agreement peers while the other did not.
After the matching, we find that between subjects who requested for second opinions at least once in the high agreement treatment and their matches, paired t-tests suggest that there are no significant differences in the values for any of the covariates. Furthermore, between the requested subjects and their matches, the standard mean differences (SMD) for most of the covariates are less or equal to 0.1, which indicates that subjects in the pairs are comparable~\cite{normand2001validating,austin2011introduction}\footnote{We have also experimented with covariate matching, and results are qualitatively similar. See the supplemental materials for details. For completeness, we also include in the supplemental materials the comparison results obtained from analyzing the raw data without applying matching methods, which are similar to results obtained after matching methods are used.}.

\begin{figure*}[t]
     \centering
     \begin{subfigure}[b]{0.24\textwidth}
         \centering
     \includegraphics[width=\textwidth]{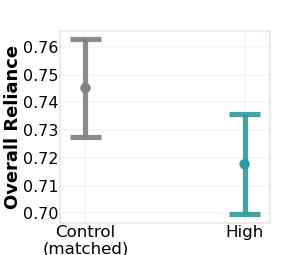}
         \caption{Overall reliance}
     \label{fig:exp2:reliance_match_high}
     \end{subfigure}
     \begin{subfigure}[b]{0.24\textwidth}
         \centering
     \includegraphics[width=\textwidth]{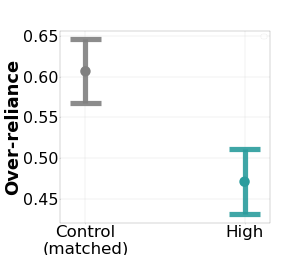}
     \caption{Over-reliance}
     \label{fig:exp2:overreliance_match_high}
     \end{subfigure}
    \begin{subfigure}[b]{0.24\textwidth}
         \centering
     \includegraphics[width=\textwidth]{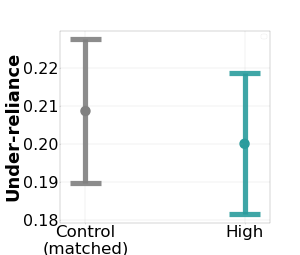}
         \caption{Under-reliance}
     \label{fig:exp2:underreliance_match_high}
     \end{subfigure}
     \centering
         \begin{subfigure}[b]{0.24\textwidth}
         \centering
     \includegraphics[width=\textwidth]{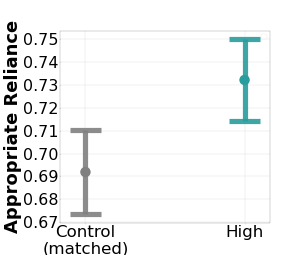}
         \caption{Appropriate reliance}
     \label{fig:exp2:approp_reliance_match_high}
     \end{subfigure}
     \vspace{-5pt}
    \caption{The effects of active solicitations of second opinions from high agreement peers on subjects' overall reliance (\ref{fig:exp2:reliance_match_high}), over-reliance (\ref{fig:exp2:overreliance_match_high}), under-reliance (\ref{fig:exp2:underreliance_match_high}), and appropriate reliance (\ref{fig:exp2:approp_reliance_match_high}) on the AI model. Data for the control treatment contains only the matched subjects after applying propensity score matching. 
    Error bars represent the standard errors of the mean.
    }
        \label{fig:exp2:reliance_match_all_high}
     \vspace{-10pt}
\end{figure*}

Figures~\ref{fig:exp2:reliance_match_high}--\ref{fig:exp2:approp_reliance_match_high} show the comparisons in subjects' overall reliance, over-reliance, under-reliance, and appropriate reliance on the AI model, respectively, between those subjects who requested for second opinions from high agreement peers for at least once and their matched subjects in the control treatment.
Conducting paired t-tests on the 31 pairs of subjects, we find that subjects' active solicitations of second opinions from high agreement peers lead to 
a significant decrease in their over-reliance on the AI model
(control: $M = 0.61, SD = 0.49$ vs. high agreement: $M = 0.47, SD = 0.50$; 
$p = 0.009$).
Importantly, we also find that the solicitation of second opinions from high agreement peers does {\em not} result in significant increases in subjects' under-reliance on the AI model. 
As a result, as shown in Figure~\ref{fig:exp2:approp_reliance_match_high}, there appears to be a slight increase in subjects' appropriate reliance on the AI model when they requested for second opinions from high agreement peers at least once (control: $M = 0.69, SD = 0.46$ vs. high agreement: $M = 0.73, SD = 0.44$), although the difference is not statistically significant at the level of $p=0.05$. 

We then repeat the propensity score matching process for the 44 subjects in the low agreement treatment who had solicited second opinions for at least once\footnote{For this matching, we added elastic (L1+L2) penalty to the logistic regression model 
to ensure the SMD in the two groups of subjects after matching are less than or equal to 0.1 on all covariates, so that the paired subjects were comparable.}, and the comparison results between the matched subjects are shown in 
Figures~\ref{fig:exp2:reliance_match_low}--\ref{fig:exp2:approp_reliance_match_low}. 
Here, we find that compared to their matched subjects in the control treatment, subjects in the low agreement treatment who solicited second opinions for at least once significantly reduced their  
overall reliance ($p < 0.001$) and over-reliance ($p = 0.003$) on the AI model, but they also exhibited significantly higher levels of under-reliance on the AI model ($p = 0.003$). Together, the active solicitations of second opinions 
from low agreement peers
does not result in a significant change in subject's appropriate reliance on the AI model. 

Together, these results show the promise of utilizing second opinions  to help people reduce their over-reliance on an AI model while not increasing their under-reliance---this goal can be achieved by enabling people to actively {\em solicit} second opinions, while in our experiment, these second opinions 
also need to 
have a relatively high level of agreement with the AI model. 
We conjecture that this approach {\em may} be effective because (1) people tend to solicit second opinions more frequently on tasks where the AI model is wrong (i.e.,  disagreements between second opinions and the AI model on these tasks lead to lower over-reliance), and (2) the relatively high level of agreement between the second opinions and the AI model minimizes the chance that people get misled by incorrect second opinions on tasks where the AI model is correct (i.e., under-reliance is not increased).

\begin{figure*}[t]
     \centering
     \begin{subfigure}[b]{0.24\textwidth}
         \centering
     \includegraphics[width=\textwidth]{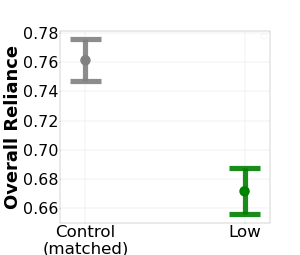}
         \caption{Overall Reliance}
     \label{fig:exp2:reliance_match_low}
     \end{subfigure}
     \begin{subfigure}[b]{0.24\textwidth}
         \centering
     \includegraphics[width=\textwidth]{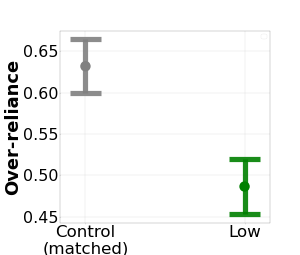}
     \caption{Over-reliance}
     \label{fig:exp2:overreliance_match_low}
     \end{subfigure}
    \begin{subfigure}[b]{0.24\textwidth}
         \centering
     \includegraphics[width=\textwidth]{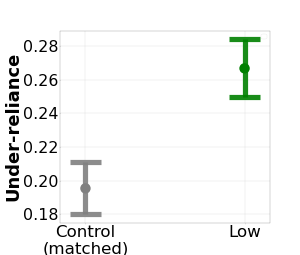}
         \caption{Under-reliance}
     \label{fig:exp2:underreliance_match_low}
     \end{subfigure}
     \centering
         \begin{subfigure}[b]{0.24\textwidth}
         \centering
     \includegraphics[width=\textwidth]{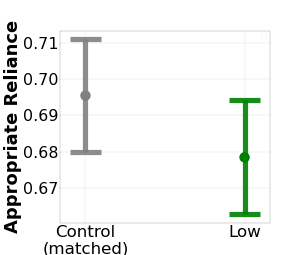}
         \caption{Appropriate Reliance}
     \label{fig:exp2:approp_reliance_match_low}
     \end{subfigure}
     \vspace{-5pt}
    \caption{The effects of active solicitations of second opinions from low agreement peers on subjects' overall reliance (\ref{fig:exp2:reliance_match_low}), over-reliance (\ref{fig:exp2:overreliance_match_low}), under-reliance (\ref{fig:exp2:underreliance_match_low}), and appropriate reliance (\ref{fig:exp2:approp_reliance_match_low}) on the AI model. Data for the control treatment contains only the matched subjects after applying propensity score matching. Error bars represent the standard errors of the mean.
    }
        \label{fig:exp2:reliance_match_all_low}
\vspace{-10pt}
\end{figure*}

\section{Discussion}

Via three randomized human-subject experiments, we investigate into how the presence of second opinions can affect people's behavior and performance in AI-assisted decision-making. 
We find that when appropriate kinds of second opinions are presented to people in appropriate formats, they are promising in helping people rely on the AI model more appropriately 
and potentially improve the decision-making performance of the human-AI team.
In this section, we discuss the
implications and limitations of our work.


\vspace{2pt}


\subsection{The Benefits, Risks, and Limitations of Second Opinions in AI-Assisted Decision-Making}
Our results of Experiment 1 show that there are clear benefits and risks when presenting second opinions from human peers to decision-makers on every single AI-assisted decision making case---it is a very effective intervention for reducing people's over-reliance on the AI model, especially when the peers have a low level of agreement with the AI model. This effect can be highly desirable in scenarios where minimizing over-reliance is the priority, such as when decisions involve high stakes. On the other hand, we also find that seeing peer-generated second opinions on every task makes decision-makers significantly increase their under-reliance on the AI model, even when the peers have a relatively high level of agreement with the AI model.  
This effect is certainly not desirable, and to the extreme, it could imply the possibility of providing ``adversarial'' second opinions to significantly reduce decision-makers' reliance on a trustworthy AI model. Overall, while we don't see that the presence of second opinions significantly changes decision-makers' appropriate reliance on the AI model/decision accuracy in Experiment 1, an interesting observation we make by comparing Figures~\ref{fig:exp1:overreliance} and~\ref{fig:exp1:underreliance} is that in all the three treatments where second opinions are presented, the magnitude of people's decrease in their over-reliance on AI is always larger than the magnitude of people's increase in their under-reliance (e.g., in the treatment with high agreement peers, over-reliance is decreased by 11.97\% and under-reliance is increased by 6.10\%). 
If this trend is generally true, it may imply that the effects of always presenting second opinions from peers to decision-makers on their decision accuracy 
depend on the AI model's accuracy---the more (less) accurate the AI model is, the more likely it will decrease (increase) the decision accuracy. 

Experiment 2 demonstrates that these double-edged effects are not specific to peer-generated second opinions. AI-generated second opinions have a similar impact on people's reliance on the primary AI recommendation. When there is a higher disagreement between the two AI models' recommendations, people tend to decrease their over-reliance on the primary model while also increasing their under-reliance. Such findings suggest that holding everything else equal, the stated source of the second opinions does not appear to significantly change how decision makers process these second opinions. This implies that when getting second opinions from human peers in the real time is infeasible, one possible alternative to consider is to use the outputs generated by another AI model to replace the human-generated second opinions. That said, we note that in our Experiment 2, in order to single out the effects of the second opinion's source on decision-makers' behavior and performance in AI-assisted decision making, we intentionally fix the content of the second opinions to be the same for treatments sharing the same level of agreement between the second opinions and the primary AI model. In practice, however, even keeping the level of agreement between the second opinions and the primary AI model the same, where human peers or the secondary AI models agree/disagree with the primary AI model may differ. Thus, future studies should look deeper into how decision-makers process second opinions when they are actually produced by real AI models. 

One possible explanation for the decreases in over-reliance and increases in under-reliance when decision-makers are presented with second opinions is that they may simply use the level of agreement between the primary AI model and the second opinion as a heuristic to gauge the primary AI model's accuracy. This accuracy estimate can then be used by decision-makers to adjust their levels of reliance on the primary AI model, without spending much effort differentiating the correctness of the AI model on individual tasks. If this is indeed the case, it may imply that the 
mere presence of second opinions from peers is not sufficient for encouraging people to engage in deep deliberative thinking on each decision-making task. To truly help people improve their decision-making performance, perhaps additional information about the second opinions (e.g., the rationale underlying the peer judgements) needs to be provided to help decision-makers make sense of them. In fact, the overly frequent presence of second opinions may even make it more convenient for people to utilize them as a heuristic in order to decrease their cognitive load rather than engage with them analytically.

In contrast, our results of Experiment 3 highlight the promise of utilizing second opinions to improve people's decision-making performance in AI-assisted decision-making by granting decision-makers the option to actively solicit second opinions, although this benefit is only observed when the second opinions 
have a relatively high level of agreement with the AI model. However, in our experiment, not every subject was willing to solicit second opinions from peers; these people may have missed the opportunity to learn from the second opinions and further improve their decision-making performance. To maximize the benefits brought about by the solicitations of second opinions, creative methods need to be designed to incentivize people to solicit second opinions or even prompt people to do so when needed.


\subsection{Design Implications for Second Opinions as an Intervention}

 {Our study provides many implications for real-world AI-assisted decision-making where the decision-maker---who is responsible for the final decision---can get ``advice'' (i.e., second opinions) from different sources. In practice, peer-generated second opinions are available in many cases. For example, a content moderator may evaluate the credibility of a social media post with the assistance of an AI-based decision aid, while they can solicit second opinions from another random member in the moderation team. 
 In these cases, second opinions can be obtained on the fly when the final decision-makers need to make their decisions. However, we note that there are many real-world scenarios, especially when a hierarchy of decision-making exists, that peer-generated second opinions will be readily available for the final decision-maker---for instance, in a security operation center (SOC), the decision recommendation of a Tier 2 analyst may already be formed before they pass on a security alert to the SOC manager for them to make the final call on how to respond. Beyond peer-generated opinions, 
 alternative AI models could also serve as a complementary source of second opinions. 
 For example, there may exist situations where peer-generated second opinions are not accessible, such as when a decision team consists of a single decision-maker with no records of historical decisions and when real-time peer consultation is not feasible due to constraints like time difference. In such scenarios, if decision-makers have access to multiple AI models, they could utilize these AI models to mimic a decision-making environment with artificial second opinions. For instance, the decision-maker could use one of the AI models (e.g., the one with the highest accuracy) as their primary AI model for AI-assisted decision making, while soliciting the ``second opinions'' from other AI models when they are unsure about the correctness of the primary model's recommendation. 
}

Based on our findings in this study, we highlight that one key premise for the provision of second opinions to effectively promote people's appropriate reliance on AI in AI-assisted decision-making is that people should {\em not} encounter too many disagreements between the second opinion and the primary AI model {\em on the tasks where the AI model is correct}. Without knowing the correctness of the AI model on each task a priori, we attempt to achieve this goal in our Experiment 3 by asking people to decide when they need second opinions and hoping that they may have a lower need for seeing second opinions on tasks where the AI is correct. Other methods can be designed to achieve this goal as well. For instance, the system may adaptively determine the presence of second opinions based on estimates of AI correctness (e.g., the AI model's confidence score). Alternatively, one may leveraging the wisdom of the crowd to present the majority opinion among a group of second advisors on each task rather than just the opinion of a randomly selected second advisor. Another important lesson from our study is whenever disagreements between the AI recommendations and the second opinions occur, support should be provided to people to help them cognitively engage with the opposing opinions and resolve the conflict by making a genuine attempt to differentiate which party is correct, instead of consuming this information in a heuristic way. To this end, other than providing the rationale for the second opinion, as we've discussed earlier, another interesting direction to explore is to combine the cognitive forcing functions---which have previously shown to be effective in nudging people to engage with the decision-making task more cognitively~\cite{buccinca2021trust}---with the presence of second opinions. In addition, knowledge about when the AI model or the second opinion can do well and when they are likely to err will also be very informative for people to decide whose recommendation to rely on when disagreements occur.


Finally, when determining whether to deploy the presence of second opinions as an intervention in AI-assisted decision making, it is also essential to consider an additional factor, that is, the cost associated with obtaining second opinions. 
For instance, when second opinions are collected from humans (e.g., peers, domain experts), it may result in financial cost to recruit them. Similarly, when second opinions are produced by AI models, it will also require training and maintaining extra AI models. Thus, whether incurring this cost to obtain second opinions is ``worthwhile'' depends on how much positive impact these second opinions can have on decision-makers' performance in AI-assisted decision making, as well as how critical making correct decisions is in the given context (i.e., the stakes of the decisions). In those cases where the introduction of second opinions brings about positive impact on one aspect of AI-assisted decision-making performance (e.g., decision accuracy) but negative impact on another aspect (e.g., decision time), one may also need to decide how to trade-off different aspects of performance.  
Ultimately, the decision on whether to incorporate second opinions as an intervention or not should be made on a case-by-case basis, weighing the potential benefits 
against the costs associated with obtaining those second opinions.

\vspace{2pt}
\subsection{On the Ecological Validity of Peer Judgements}
As discussed in Section~\ref{sec:design}, in this study, to ensure the ecological validity of the peer-generated second opinions, we collected them from real crowd workers using a pilot study. We note that these real-world peer judgements have some important characteristics that are likely critical for them to be useful for promoting people's appropriate reliance on AI. For example, for all three sets of peers we've created in our study, their average level of agreement with the AI model is higher when the AI is correct than when the AI is wrong (AI correct: 82.22\%, 51.11\%, and 31.11\% for high, medium, low agreement peers, respectively; AI incorrect: 60.00\%, 46.67\%, and 26.67\% for high, medium, low agreement peers, respectively). This characteristic indirectly ``help'' decision-makers encounter fewer peer-AI disagreements on tasks where the AI model is correct than what would have been observed if the peers disagree with the AI equally frequently regardless of AI correctness or even disagree with the AI more when the AI is correct. In fact, despite our key finding in Experiment 3 is that the active solicitations of second opinions may mitigate over-reliance on AI without increasing under-reliance if peer judgements have a relatively high level of agreement with AI, we suspect that not all ``high level of agreement with AI'' is created equal---when fixing the level of agreement between peers and the AI model, peers that fully agree with AI when it is wrong but have some disagreements with it when it is correct is unlikely to help promote decision-makers' appropriate reliance on AI. However, with great ecological validity comes great challenges in isolating the effects of different factors. For example, in our study, as we increase the level of agreement between real-world peers and the AI model from low to high, not only do the peers agree with the AI model's recommendations more frequently, but their own accuracy is increased while their errors become less independent of those of the AI model's. Carefully separating how each of these factors of the peer-generated second opinions, alone, affect decision-makers' behavior and performance in AI-assisted decision-making will be a very interesting and important future direction.

\vspace{2pt}
\subsection{Limitations and future work} Our study was conducted with laypeople (i.e., subjects recruited from MTurk) on a decision-making task that does not require much expertise yet still seems to be not easy for laypeople (e.g., in our pilot study, the average decision accuracy of the crowd workers on our selected sentiment analysis task is 64.34\%). Cautions should be used when generalizing the results of this work to different settings, such as for a different population of people, for tasks that are much easier, or for tasks that require substantially more domain expertise. 

In addition, the second opinion in our experiment came from randomly selected crowd workers that our subjects did not know. In other words, our experiment reflects a scenario where decision-makers can request a ``system'' to obtain second opinions for them in AI-assisted decision making, while decision-makers themselves are not directly involved in the process of identifying who to solicit the second opinions from. In reality, decision-makers in AI-assisted decision making 
may actively seek help and solicit second opinions from those people that they are quite familiar with and naturally trust, which may significantly change how they respond to the agreements or disagreements between the AI recommendations and the peer's second opinions. Future studies should be conducted to understand when decision-makers actively involve in identifying their additional advisors to seek second opinions from, how those second opinions will affect their behavior and performance in AI-assisted decision making. 

We believe our experiment setup (e.g., the choice of sentiment analysis tasks, the usage of MTurk workers as our human subjects) could well represent some specific AI-assisted decision making scenarios (e.g., AI-assisted data labeling). For instance, we found that subjects in our experiment spent a very short amount of time on each decision making task (about 5--8 seconds on average) and some of them never solicit second opinions when given this option. This behavior may be attributed to crowd workers' nature of optimizing for the speed, but it reflects real-world annotators' behavior in AI-assisted data labeling well. 
That said, we acknowledge that our experimental setup does not reflect all different kinds of AI-assisted decision making settings. For example, our experimental setup may not be representative of those decision-making scenarios involving high stakes, in which decision-makers will likely
spend more time, exhibit stronger motivations, and engage in more analytical thinking on each decision making case. Also, as discussed earlier, second opinions that are generated by real AI models may possess different characteristics from the ones generated by humans, and future studies should investigate into their impacts in more depth. 

Furthermore, 
we note that the structure of the advice in AI-assisted decision making is also not limited to the combination of one single primary recommendation by the AI model and one second opinion, as studied in this work. For example, 
multiple second opinions can be provided instead of one, and the second opinions may also come from a combination of sources (e.g., laypeople, domain experts, AI) rather than a single source. 
Understanding how second opinions under these settings affect decision-makers' behavior and performance in AI-assisted decision-making is another exciting future work. Finally, interesting future work could be carried out to delve deeper into whether there exist any individual differences on the effects of second opinions in AI-assisted decision making, and how these effects may evolve over time.


\section{Conclusion}

In this paper, we explore the effect of providing second opinions to people on their behavior and performance in AI-assisted decision-making. Via three pre-registered, randomized experiments, we show that always presenting second opinions along with the AI recommendation can reduce decision-makers' over-reliance on AI and increase their confidence in their correct decisions, but it also increases decision-makers' under-reliance on AI. Such effects hold regardless of whether the second opinions are provided by human peers or another AI model. Nevertheless, 
by enabling decision-makers to actively solicit second opinions from peers as needed, we find that decision-makers' active solicitations of second opinions have the promise to reduce their over-reliance on the AI model without increasing the under-reliance in some cases.  
Our results highlight the potential benefits, risks, limitations, and implications of presenting second opinions to people in AI-assisted decision making for promoting the human-AI team performance. We hope this work could open more discussions on understanding the effects of second opinions in AI-assisted decision-making and better utilizing them as an intervention to enhance human-AI collaboration in decision-making. 

\bibliographystyle{ACM-Reference-Format}
\bibliography{peer}


\end{document}